%% file: dissertation.tex
\documentclass[dissertation]{uathesis3}

\usepackage{graphicx}
\usepackage{pdfpages}
\usepackage[square,numbers,sort&compress]{natbib}
\usepackage[titles]{tocloft}

\newcommand{\doi}[1]{\href{http://dx.doi.org/#1}{DOI:#1}}

\usepackage{amsmath,amssymb}			
\usepackage[pdftex,bookmarks,colorlinks=true,urlcolor=black,linkcolor=black,citecolor=black]{hyperref}
\usepackage{float}
\usepackage{wrapfig}

\providecommand{\sfrac}[2]{\ensuremath{{}^{#1}\!/ \:\!\!{}_{#2}}}

\completetitle{Charm and strangeness in quark-gluon plasma hadronization}
\fullname{Michal Petran}			
\degreename{Doctor of Philosophy}	

\begin{document}
\newlength{\mylength}

\maketitlepage
{DEPARTMENT OF PHYSICS}	
{2013}							

\approval
{5 November 2013}		
{Johann Rafelski}	
{Johann Rafelski}  
{Sean Fleming}		
{Ken Johns}			
{Michael Shupe}		
{Koen Visscher}		

\statementbyauthor


\incdedication{dedication}

\tableofcontents

\listoffigures


\incabstract{abstract}

\include{00_introduction}

\include{01_PresentStudy}

\include{02_Summary}

\appendix
\include{A00_XiOverPhi}
\include{A01_RHIC62}
\include{A02_UniversalHadronization}
\include{A03_Alice2.76TeV}
\include{A04_SQM2013-strangeness}
\include{A06_Share3Manual}
\include{A05_SQM2013-charm}

\renewcommand{\baselinestretch}{1}		
\small\normalsize						

\bibliographystyle{MyBib} 
\addcontentsline{toc}{chapter}{REFERENCES}
\bibliography{bibliography}

\renewcommand{\baselinestretch}{1.4}		
\small\normalsize		

\end{document}

%% file: 00_introduction.tex
\chapter{QUARK-GLUON PLASMA IN THE LABORATORY}
\label{chap:intro}
\section{Particle production in heavy-ion collisions}

\subsection{Quark-gluon plasma -- a new phase of matter}

In matter around us, only colorless combinations of strongly interacting elementary particles, quarks and gluons, are present. In the Early Universe, however, all matter was sufficiently hot and dense so that quarks and gluons were free and formed a new phase of matter we refer to as `quark-gluon plasma' (QGP). The QGP has been under active investigation in past decades~\citep{Jacak:2012dx}. Studying QGP in laboratory experiments aims to answer fundamental questions about the structure of matter. This requires exploration of collective phenomena of the strongly interacting medium and understanding of the transition between this new QGP phase of matter and the regular matter that emerges in a process called hadronization.

In heavy-ion collisions at relativistic energies, at first a dense partonic matter is created. Such strongly interacting matter at high enough energy density dissolves into a thermal gas of quarks and gluons, the state we refer to as quark-gluon plasma. When left to expand and cool down, the quarks bind into colorless combinations of three quarks ($qqq$) called baryons and quark-antiquark ($q\overline{q}$) states called mesons. The process the quark-gluon plasma undergoes while freezing-out the color degree of freedom is called `hadronization'. It has been of particular interest to many for several decades and is to this dissertation as well. The description of hadronization in strong coupling limit leads to phase space dominance model for particle production. This statistical hadronization model (SHM) has been successfully used to describe hadron production in high energy particle collisions, in particular heavy-ion 
\begin{wrapfigure}{l}{0.43\textwidth}
\centering
\includegraphics[height=0.88\textheight]{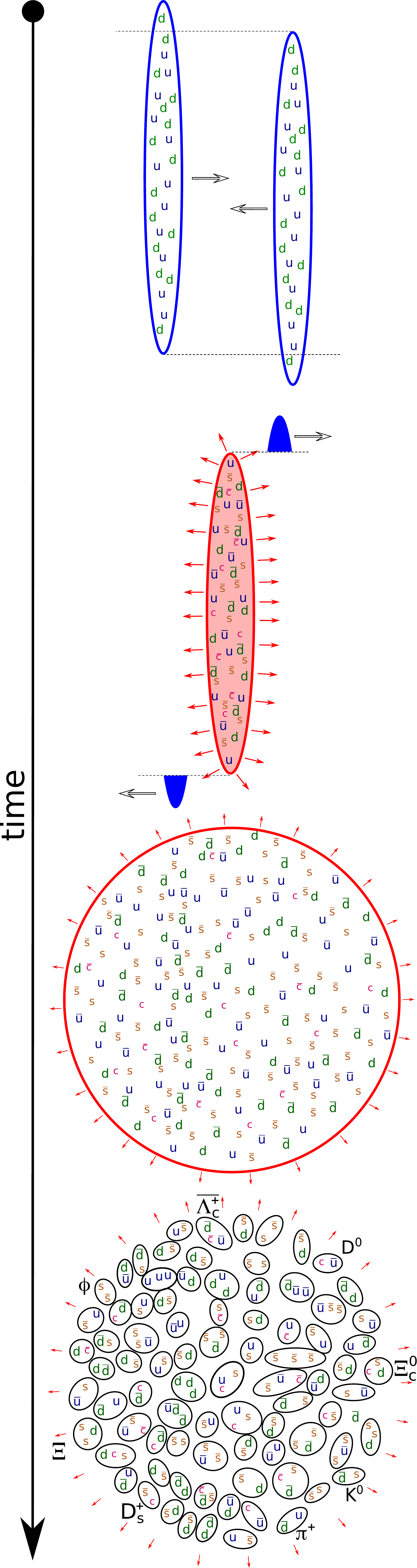}
\hfill
\caption[Heavy-ion collision evolution]{\label{fig:HI}Schematic heavy-ion collision evolution.}
\vspace*{-10pt}
\end{wrapfigure}
collisions, in conditions where we expect the creation of quark-gluon plasma.

The main stages of such collisions are depicted in Figure~\ref{fig:HI}. In the laboratory frame of a collider experiment, the Lorentz contracted ions collide. The initial hard parton scattering is the dominant mechanism of heavy ($c,b,t$) quark production. A droplet of QGP forms, thermalizes and expands. During the expansion and cool-down, light quarks ($u,d,s$) are produced abundantly, and heavy quarks are carried by the medium outwards from the collision center. When the conditions of the QGP fireball reach a critical point, the fireball hadronizes. After this point, quarks are present only in newly produced bound states, hadrons.

The confinement of quarks and gluons in hadrons is a phenomenon investigated in the context of the theory of strong interactions, quantum chromo-dynamics (QCD). It is not the topic of this dissertation; however, quarks and gluons interact in the QGP. Because the coupling constant of QCD, $\alpha_s$, changes as a function of energy scale in the interactions, only at high momentum transfer ($\gg 1$~GeV) is it small enough to allow the use of perturbative theory methods. At low momentum transfer where hadronization takes place, the perturbative expansion does not apply. In the limit of strong coupling, we have to rely on phenomenological models, such as SHM. SHM enables us to study and  understand the properties of the QGP fireball, the hadronization conditions at which the color degree of freedom freezes-out, and the connection between the chemistry of the plasma and emerging hadrons arises.

\subsection{Statistical hadronization model}
\label{sec:SHM}
The statistical hadronization model aims to describe hadron production in heavy-ion collisions given thermal parameters of the particle source, whether or not deconfinement has been achieved. Given a set of common thermal parameters, hadronization temperature $T$, system volume $V$ and chemical potentials $\mu$, the phase space of all hadrons is populated. Below we present the essence of SHM, showing the necessary steps to calculate the hadron yields using statistical methods. More detailed description is presented in Section 2 of Appendix~\ref{apx:ShareManual} and a thorough derivation can be found in~\citep{Letessier:2002gp}.

We work in grand-canonical limit when describing the phase space of produced particles. Each hadron species $i$ is thus described in SHM by the distribution
\begin{equation}
\label{eq:distribution}
n_i\equiv n_i\left( E_i \right) =\frac{1}{\Upsilon_i^{-1}\exp\left( E_i/T \right)+S},\quad 
S=\left\{
\begin{array}{ll}
+1 & \; \text{Fermi-Dirac distribution} \\[-2mm]
0 & \; \text{Boltzmann approximation}\\[-2mm]
-1 & \; \text{Bose-Einstein distribution.} \\
\end{array}
\right.
\end{equation}
where the hadron energy $E_i = \sqrt{m_i^2+p_i^2}$, $T$ is the temperature of the source, $\Upsilon_i$ is the fugacity and $S$ distinguishes the appropriate distribution for fermions, bosons and Boltzmann approximation.

The introduction of the temperature $T$ of the source means that we assume produced particles to be in kinetic, often called `thermal', equilibrium, but not necessarily in chemical equilibrium. The fugacity $\Upsilon_i$ describes the chemical potential of the $i$-th hadron based on its constituent quark content. The total fugacity of a hadron $i$ is a product of constituent quark ($u,d,s,c$) fugacities,
\begin{equation}
\Upsilon_i = \prod_{f=u,d,s,c} \Upsilon_f.
\end{equation}
Fugacity of each quark flavor $\Upsilon_f$ is usually further decomposed into a product of two factors, the phase space occupancy $\gamma_f$, and chemical potential factor $\lambda_f$.
\begin{equation}
\Upsilon_f = \gamma_f \lambda_f.
\end{equation}
The fugacity factor $\lambda$ relates to quark chemical potential $\mu_f$ of each quark flavor $f$ as
\begin{equation}
\lambda_f = e^{\mu_f/T}
\end{equation}
It can be shown that for anti-particles (anti-quarks), $\gamma_{\bar{f}}=\gamma_f$ and $\lambda_{\bar{f}}=\lambda_f^{-1}$~\citep{Letessier:2002gp}.

The simplest and most commonly used SHM approach assumes that all hadron phase space populations reach an equilibrium abundance, which is equivalent to fixing all $\gamma$'s to unity. The introduction of phase space occupancies $\gamma_f\neq 1$ originates from the recognition that it takes considerable time to re-equilibrate flavor abundance. In the scenario of very dynamical, sudden QGP hadronization~\citep{Rafelski:2000by}, the number of available quarks of each flavor to be distributed among final state hadrons is given by the conditions in QGP at hadronization. Bearing in mind that equilibrium abundance in the quark phase does not have to be the same as chemical equilibrium abundance of valence quarks in the hadrons, we have to allow for the chemical non-equilibrium. Note that if the hadron abundances were indeed in absolute chemical equilibrium, this general approach allows the phase space occupancies $\gamma_f$ to converge to unity. As we will show in Chapter~\ref{chap:presentstudy}, this is not the case.

\begin{figure}[t]
\centering
\includegraphics[width=0.55\columnwidth]{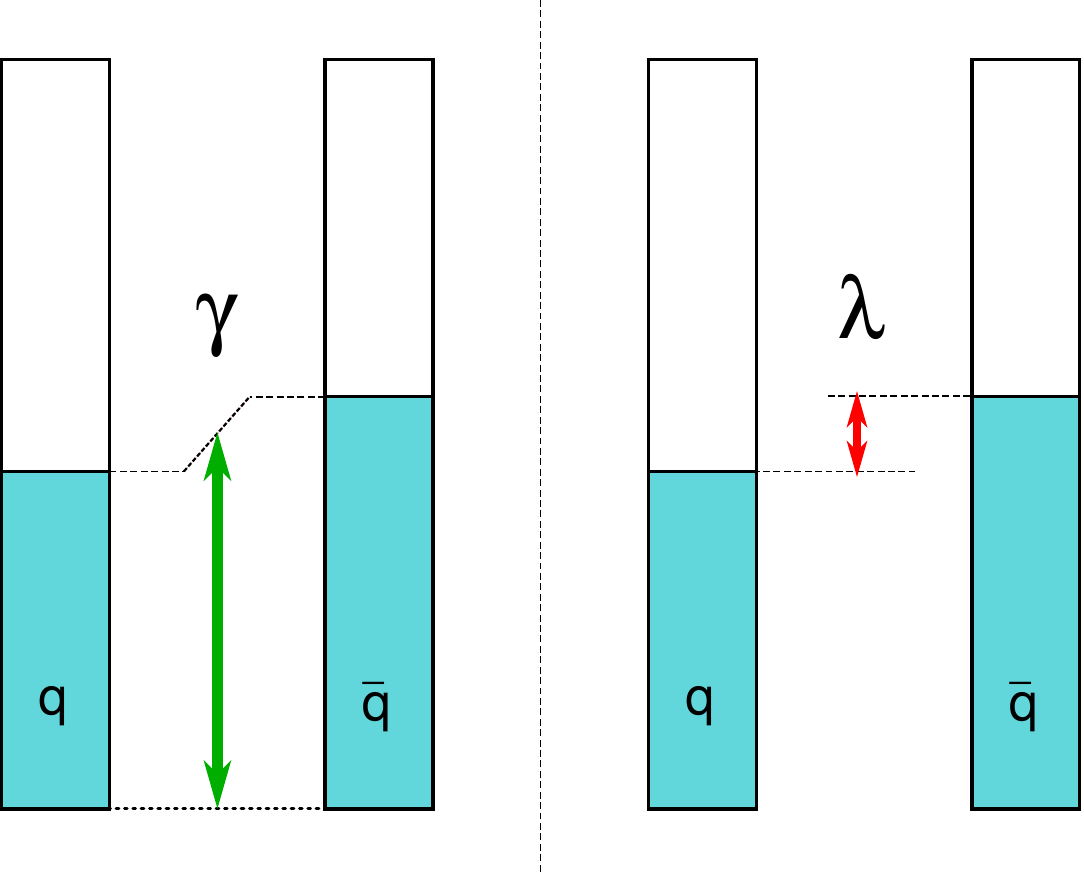}
\caption[The effect of phase space occupancy and fugacity factor]{\label{fig:ChemReactions} Illustration of $\gamma$ and $\lambda$ controlling quark abundance. The phase space occupancy $\gamma$ controls the population of quarks $q$ and antiquarks $\bar{q}$ (left); the fugacity factor $\lambda$ controls the $q-\bar{q}$ difference (right).}
\end{figure}

The parameters $\gamma_f$ controlling the number of $q_f,q_{\bar{f}}$ quark pairs of flavor $f$, as we just explained, is responsible for what one calls, \emph{absolute} chemical equilibrium. The parameter $\lambda_f$ describes the \emph{relative} chemical equilibrium, that is the difference of quarks and anti-quarks $q_f-q_{\bar{f}}$ in the system. The effects of $\lambda$ and $\gamma$ are schematically depicted in Figure~\ref{fig:ChemReactions}. The phase space occupancy $\gamma$ describes the number of quark-antiquark pairs present in the system, whereas $\lambda$ describes the difference between the quarks and antiquarks of the same flavor~\citep{Koch:1986ud}.

For a hadron $i$ composed of $N_u, N_d ,N_s$ and $N_c$ up, down, strange and charm quarks and $N_{\bar{u}},N_{\bar{d}},N_{\bar{s}}$ and $N_{\bar{c}}$ anti-quarks respectively, the fugacity can be expressed as:
\begin{equation}
\label{eq:fugacity}
\Upsilon_i = 
(\gamma_u^{N_u+N_{\bar{u}}} \lambda_u^{N_u-N_{\bar{u}}}) \, 
(\gamma_d^{N_d+N_{\bar{d}}} \lambda_d^{N_d-N_{\bar{d}}}) \,
(\gamma_s^{N_s+N_{\bar{s}}} \lambda_s^{N_s-N_{\bar{s}}}) \,
(\gamma_c^{N_c+N_{\bar{c}}} \lambda_c^{N_c-N_{\bar{c}}})
\end{equation}

Given the temperature $T$ and fugacities of all flavors, the hadron yield in a given volume $V$ can be calculated by integrating the distribution (Eq.\ref{eq:distribution}) over the phase space:
\begin{equation}
\label{eq:yield}
\langle N_i \rangle = g_i V \int\frac{\mathrm{d}^3p}{(2\pi)^3} \, n_i,
\end{equation}
where $g_i$ is the hadron spin degeneracy $g=(2J+1)$. For a massive particle, Eq.\ref{eq:yield} can be numerically evaluated to any desired precision when expressed as a series of modified Bessel functions of the second kind
\begin{equation}
\langle N_i \rangle = \frac{g_i V T^3}{2\pi^2}\sum\limits_{n=1}^\infty\frac{(\pm 1)^{n-1}\Upsilon_i^n}{n^3}\left(\frac{nm_i}{T}\right)^2{K_2}\left(\frac{nm_i}{T}\right).\label{eq:yieldexpansion}
\end{equation}
This is widely used in SHM implementations including the program we developed, SHARE, which I upgraded with a CHARM calculation module, and tuned it for LHC (see Appendix~\ref{apx:ShareManual}), and which was used to produce the results presented.

As discussed so far, the observation of all particles of type $i$ is required to determine the total volume $V$ at hadronization. However, we work with RHIC and LHC results, which were obtained in mid-rapidity range. Thus we study ${d\langle N_i\rangle}/{dy}$ and accordingly determine a volume parameter ${dV}/{dy}$. This is a 3-dimensional freeze-out hypersurface of the 4-dimensional domain generating particles within $-0.5 < y < 0.5$ domain. This dissertation does not relate ${dV}/{dy}$ to a geometric and/or dynamical interpretation.

Once we obtain a set of statistical parameters $dV/dy, T, \gamma_i$, and $\lambda_i$ from a best fit to experimental data, we can evaluate within SHM all and any hadron yield and the physical properties of the QGP at hadronization, the latter by summing the contributions of all emerging hadrons carrying away the energy, entropy, and other properties.

\section{Strangeness and charm production and freeze-out}
\label{sec:intro-hadronization}
\subsection{Origin of strangeness and charm}
The colliding nuclei contain only a very small fraction of the second family quarks, strange or charm. The dominant production mechanisms of these two flavors in relativistic heavy-ion collisions are qualitatively different. Both flavors are produced in pairs $s+\bar{s}$ and $c+\bar{c}$ in the early stage of a heavy-ion collision in hard parton scattering. Upon thermalization, the initial fireball temperature is of the order of $300-600$ MeV~\citep{Jacak:2012dx}, rapidly decreasing during the fireball expansion. There is a qualitative difference in the subsequent behavior of strange and charm quarks, because during the evolution of the QGP fireball, its temperature falls between the strange and charm quark mass $m_s < T^{\small QGP} < m_c$, where $T = 140\,\mathrm{MeV} < T^{\small QGP} < T_{initial}^{\small QGP} \sim 600\,\mathrm{MeV}$.
\begin{enumerate}
\item
The strange quark mass of $m_s = 95\pm5\,\mathrm{MeV}/c^2$~\citep{Beringer:1900zz} is lower than the QGP temperature. This allows a significant additional amount of strangeness to be produced in thermal gluon fusion processes during the expansion of the fireball, which greatly enhances the total yield of strangeness available at hadronization~\citep{Rafelski:1982pu}. QGP can achieve chemical equilibrium, which implies $\gamma_s>1$ in the hadron phase.
\item
Due to the high mass of charm flavor compared to the QGP laboratory temperature , $m_c = 1275\pm25\,\mathrm{MeV}/c^2$~\citep{Beringer:1900zz}, virtually all charm is produced in the early hard parton scattering processes. Depending on the total amount of charm produced, or rather its density in the fireball, it can undergo re-annihilation during the fireball expansion, as long as $\gamma_c>1$. In general, this is a slow process compared to strangeness. We cannot expect charm to be in absolute chemical equilibrium in QGP, and hence not in the hadron phase either. It is therefore necessary to introduce the charm phase space occupancy factor $\gamma_c$ which we expect to be $\gamma_c\gg 1$ at hadronization.
\end{enumerate}
One of the objectives of this dissertation is to quantify the effect of charm hadrons production and decay on the final measured particle abundances and hence we focus on the amount of charm present at the end of the fireball expansion, at hadronization. We use as a measure of charm abundance the total charm, i.e., the number of charm and anticharm quarks $N_{c\bar{c}}= (c+\bar{c})$ present at hadronization. Although the charm is described by the equations introduced in Section~\ref{sec:SHM} the same way as the other flavors with $\gamma_c$ and $\lambda_c$ as parameters, it is more intuitive to evaluate $\gamma_c$ given the value of $N_{c\bar{c}}$, which thus becomes a model parameter.

There are two reasons to allow a separate charm hadronization temperature $T_{\mathrm{charm}}$:
\begin{enumerate}
\item Charm quarks may not completely follow the kinetic equilibrium within QGP in the first pre-hadronization stage.
\item Due to their relatively large binding, charm hadrons can preform earlier in the QGP and thus the final abundances is controlled by a higher chemical freeze-out temperature $T_{\mathrm{charm}}$.
\end{enumerate}

\subsection{Statistical hadronization of charm}
In the statistical model approach, the number of charm (and anticharm) quarks is distributed among the charm hadron states according to the relative abundance of other flavors present. Following the approach of~\citep{Kuznetsova:2006bh}, given $N_{c\bar{c}}$ and all other model parameters, we solve for charm phase space occupancy $\gamma_c$
\begin{align}
\label{eq:gamma_c}
\langle N_{c\bar{c}} \rangle &= \gamma_c\left(\gamma_q \langle N^{eq}_{qc} \rangle + \gamma_s \langle N^{eq}_{sc} \rangle
+ \gamma_q^2\langle N^{eq}_{qqc} \rangle + \gamma_s\gamma_q\langle N^{eq}_{cqs} \rangle + \gamma_s^2\langle N^{eq}_{ssc} \rangle
\right)\nonumber\\
&+ \gamma_c^2 \left(\langle N^{eq}_{cc} \rangle
+ \gamma_q\langle N^{eq}_{ccq} \rangle
+ \gamma_s\langle N^{eq}_{ccs} \rangle \right)\nonumber\\
&+ \gamma_c^3\langle N^{eq}_{ccc} \rangle,
\end{align}
where $\langle N^{eq}_{12(3)} \rangle$ is the statistical equilibrium yield of a meson with quark content `$12$' or a baryon with quark content `$123$'.

\begin{figure}[t]
\centering
\includegraphics[scale=0.85]{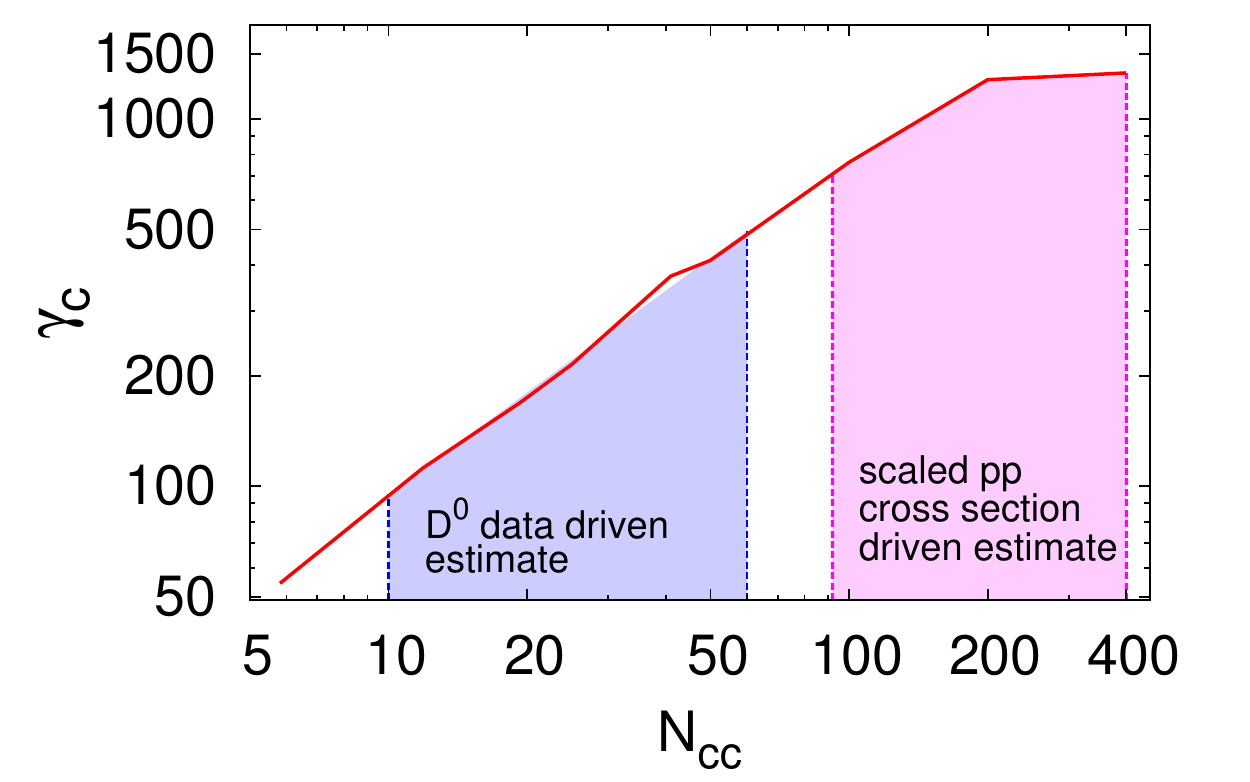}
\vspace*{-10pt}
\caption[Charm phase space occupancy as a function of total charm $N_{c\bar{c}}$]{\label{fig:gammac}Charm phase space occupancy as a function of total charm $N_{c\bar{c}}$ present at hadronization in 0--20\% centrality Pb-Pb collisions at $\sqrt{s_{NN}}=2.76\,\mathrm{TeV}$. In this example, $\gamma_c$ is computed using $T_{\mathrm{charm}}=T$.}
\end{figure}

As Figure~\ref{fig:gammac} shows, the charm phase space occupancy $\gamma_c$ increases very rapidly as $\gamma_c \simeq 16 N_{c\bar{c}}^{\,0.8}$, reaching $\gamma_c=100-500$ for the $D^0$ meson data inspired range of $N_{c\bar{c}}$ and $\gamma_c >1000$ for the scaling charm pair production cross section estimate.

\subsection{Estimated charm abundance at RHIC and LHC}
\label{sec:charm-estimates}
\begin{figure}[t]
\centering
\includegraphics[scale=0.8]{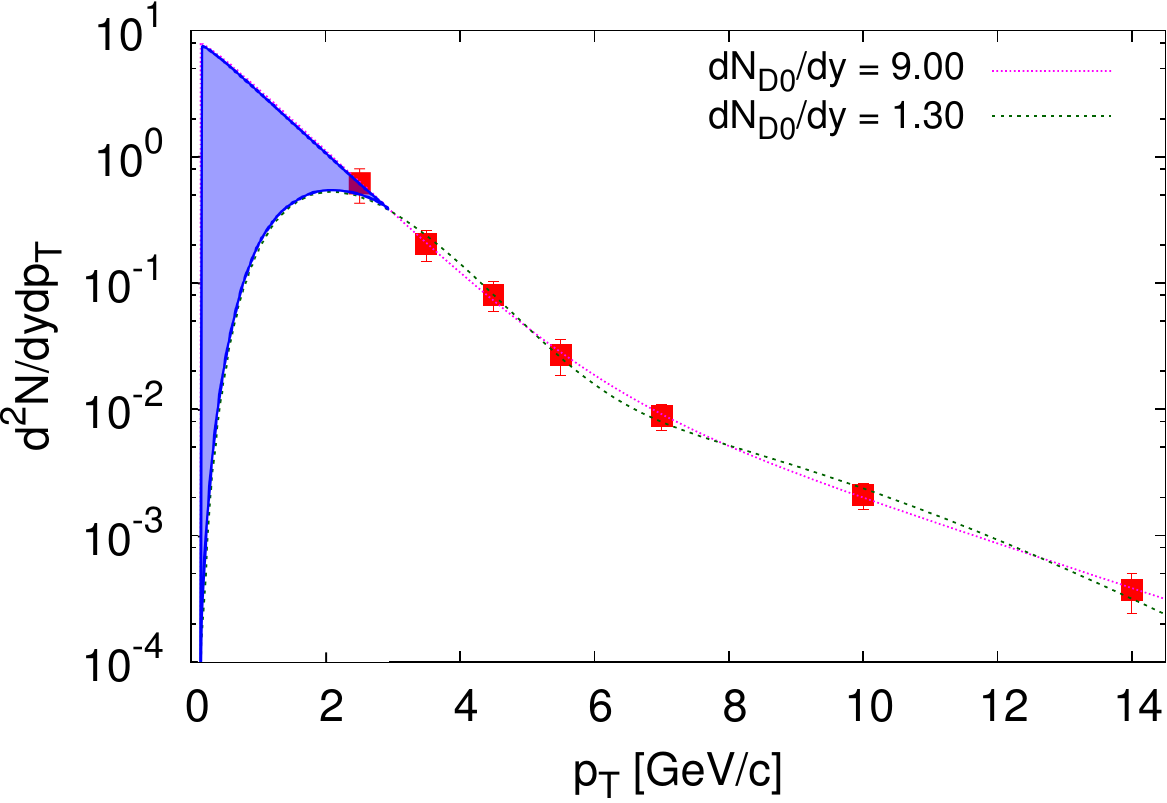}
\caption[$D^0$ meson $p_\perp$-spectrum from central Pb--Pb collisions at LHC]{\label{fig:DZero} Shaded area shows a domain of extrapolation of the $D^0$ meson $p_\perp$-spectrum from central Pb--Pb collisions at $\sqrt{s_{NN}}=2.76\,\mathrm{TeV}$, data points measured by the ALICE experiment~\citep{ALICE:2012ab}.}
\end{figure}

At the time of this dissertation preparation, the total charm cross-section of $d\sigma_{c\bar{c}}/dy|_{y=0}=175\pm12\pm23\,\mu b$ has been measured at RHIC and has been shown to scale with the number of binary collisions~\citep{Tlusty:2012ix}. Multiplying the cross-section with the number of binary collisions in a central Au--Au collision at 200 GeV ($N_{\mathrm{bin}}= 1012\pm59$~\citep{Abelev:2008ab}) and normalizing by the total inelastic cross-section in $pp$ collisions, the estimated total charm abundance is
\begin{equation}
\label{eq:Ncc-RHIC}
N^{\mathrm{RHIC}}_{c\bar{c}} = N_{\mathrm{bin}}\frac{d\sigma^{NN}_{cc}/dy}{d\sigma^{pp}_{\mathrm{total}}/dy} \simeq 8.6.
\end{equation}
Based on the statistical hadronization of charm with $T_{\mathrm{charm}}=T$, the theoretical $D^0$ meson yield we obtain is compatible with the experimental data $dN_{D^0}/dy = 3.4 \pm 0.32\,\mathrm{(stat.)} \pm 0.28\,\mathrm{(syst.)}$~\citep{Yifei:2013zz} for 0--10\% centrality Au-Au $200\,\mathrm{GeV}$ collisions, .

No charm hadron yield has yet been measured in heavy-ion collisions at LHC. An estimate may be based on the incomplete $p_\perp$-spectrum of $D^0$ meson~\citep{ALICE:2012ab}, see Figure~\ref{fig:DZero}, where the low $p_\perp$ bins are yet to be resolved and the spectrum extrapolated to $p_\perp=0$. We extrapolate the spectrum without the missing points to estimate a range of $D^0$ yield $dN_{D^0}/dy \in (1.3,9.0)$. Considering the statistical hadronization of charm, this corresponds to $N^{\mathrm{LHC}}_{c\bar{c}} \in (6,45)$, see corresponding domain of $\gamma_c$ in Figure~\ref{fig:gammac}. From the charm production cross section measurement in $pp$ collisions, the estimated amount of charm produced in a central Pb--Pb collision at LHC is much higher, $N_{c\bar{c}} = 246\pm154$, where the large uncertainty originates in the uncertainty of the total charm cross section. The relevant domain of $\gamma_c$ is shown in Figure~\ref{fig:gammac}. It is important to remember that the charm abundance estimate based on $D^0$ yield is directly related to the charm yield at hadronization, whereas the theoretical based calculation reflects the amount of charm produced in the initial stages of the collision. The discrepancy between the two may be due to the annihilation of $c\bar{c}$ pairs during the fireball expansion.

%% file: 01_PresentStudy.tex
\chapter{PHYSICS OF STRANGENESS AND CHARM HADRONIZATION}
\label{chap:presentstudy}
This chapter summarizes results that are presented in detail in the appended publications, and which are referenced as appropriate throughout this chapter. The presentation follows the time line of my research work and its publication of the past years.

\section{Hadron ratios as flavor probes}
\label{sec:ratios}
The QGP flavor content is imprinted on the hadron distribution. Strangeness rich QGP leads to the production of more strange and even more multistrange particles.  One can describe all measured hadron yields fitting them with a complex multi-parameter SHM model. However, as a first step, we choose a different simplified approach (see Appendix~\ref{apx:MultistrangeRatios}), in which we explore hadron production at RHIC at $\sqrt{s_{NN}}=62.4\,\mathrm{GeV}$ testing in the process the consistency of SHM. Consideration of specific ratios of hadrons rather than their individual yields removes the dependence on most model parameters. Consider the following ratio:
\begin{align}
\label{eq:XiPhi}
\frac{\Xi}{\phi} &\equiv \sqrt{\frac{\Xi^-(ssd)\,\overline{\Xi}^+(\bar{s}\bar{s}\bar{d})}{\phi(s\bar{s})\;\phi(s\bar{s})}} = \sqrt{\frac{\gamma_s^4\gamma_q^2}{\gamma_s^4}
\frac{\lambda_s^2\lambda_s^{-2}\lambda_q\lambda_q^{-1}}{\lambda_s^2\lambda_s^{-2}}} 
\left( \frac{dV_\Xi}{dy} \right) \left( \frac{dV_\phi}{dy} \right)^{-1} f(T,m_\Xi,m_\phi)\nonumber\\
 & = \gamma_q f(T,m_\Xi,m_\phi).
\end{align}
In this form, by taking a ratio of the same number of hadrons in the denominator and in the numerator, we remove the dependence on overall normalization; in Eq.~\ref{eq:XiPhi}, the overall normalization is volume per unit rapidity $dV/dy$. Because both hadron species originate from one QGP fireball source, $dV_\Xi/dy = dV_\phi/dy$. Multiplying a particle with its antiparticle causes the fugacity factors (or equivalently the chemical potentials) to cancel. Finally, by choosing specific hadron species, we can eliminate dependence on different powers of phase space occupancies; in Eq.~\ref{eq:XiPhi}, the dependence on $\gamma_s$ vanishes. The $\Xi/\phi$ ratio is then directly proportional to $\gamma_q$ and a function of the hadronization temperature $T$ and the mass of $\Xi$ and $\phi$. Note that the yield of $\Xi$ is a sum of the $\Xi$ and higher mass resonances with the same quark content with $\Xi(1530)^-$ as the second most significant contribution. Looking at this ratio as a function of reaction energy and centrality enables us to study primarily the model parameter, $\gamma_q$, and to some degree $T$, independently of the other parameters, $dV/dy, \gamma_s, \lambda_s, \lambda_q$, as is indicated in the second form of Eq.~\ref{eq:XiPhi}.

\begin{figure}
\centering
\includegraphics[scale=0.9]{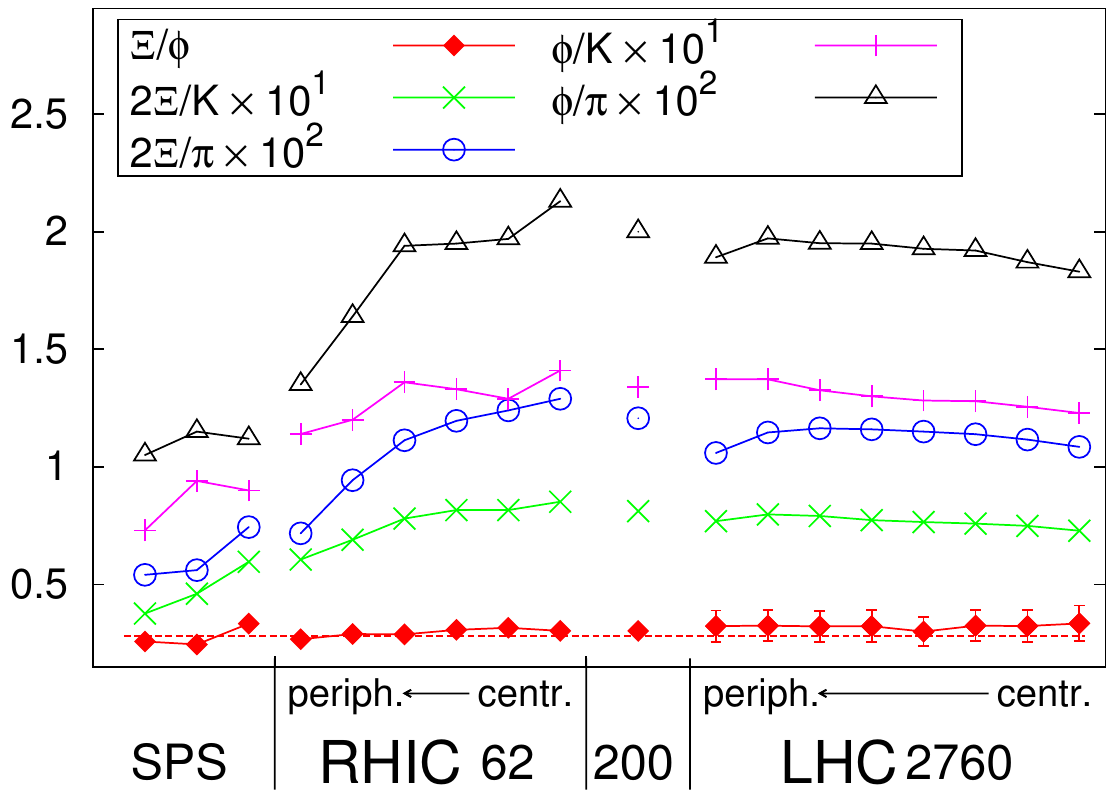}
\caption[Hadron ratios from SPS to LHC]{\label{fig:XiPhi-ratio}Hadron ratios at different heavy-ion experiments from SPS to LHC.}
\end{figure}

Data shown in Figure~\ref{fig:XiPhi-ratio} from various heavy-ion experiments ranging from GeV to TeV energies demonstrate that $\Xi/\phi$ as defined by Eq.~\ref{eq:XiPhi} is constant over the whole range of energy, and at different centralities. This implies that $\gamma_q$ and $T$ describing all these data sets will be similar. The remarkable universality of the $\Xi/\phi$ ratio also validates the SHM model.  A different result is obtained from the study of ratios proportional to $\gamma_s$ which change significantly even at one energy, e.g., 62.4 GeV, but different centralities. This implies the need for strangeness phase space occupancy $\gamma_s$ as a strongly variable model parameter. This result is obtained by considering the ratio of 
\begin{equation}
\label{eq:Xi-K}
\Xi/K \equiv \sqrt{\frac{\Xi^-(ssd)\,\overline{\Xi}^+(\bar{s}\bar{s}\bar{d})}{K^+(u\bar{s})\;K^-(s\bar{u})}} \propto \gamma_s,
\end{equation}
which increases by 50\%  for RHIC 62.4 GeV as a function of centrality. Even more pronounced functional dependence (factor of 2) is observed in Figure~\ref{fig:XiPhi-ratio} for ratios proportional to $\gamma_s^2$, such as 
\begin{equation}
\label{eq:Xi-pi}
\Xi/\pi \equiv \sqrt{\frac{\Xi^-(ssd)\,\overline{\Xi}^+(\bar{s}\bar{s}\bar{d})}{\pi^+(u\bar{d})\;\pi^-(d\bar{u})}} \propto \frac{\gamma_s^2}{\gamma_q},
\end{equation}
where $\gamma_q$ is fixed to a common value by the constant $\Xi/\phi$ ratio. This implies that strangeness reaches different levels of equilibration depending on the available collision energy and the reaction volume size (and therefore lifespan) of the QGP fireball. With a clear need for $\gamma_s$ for proper description of hadron production in heavy-ion collisions, we effectively ruled out the chemical equilibrium model with all $\gamma$'s fixed to unity. We established not only on a theoretical, but now also on an experimental basis, that strangeness is not chemically equilibrated in the observed hadron yields. This result confirms and extends the first analysis of relativistic heavy-ion collision experimental data based on the SHM principles~\citep{Rafelski:1991rh}.

\section{Strangeness overabundance and hadronization conditions at RHIC}
\label{sec:StrangenessAtRHIC62}
Considering all the available data from Au--Au collisions at $\sqrt{s_{NN}}=62.4\,\mathrm{GeV}$~\citep{Aggarwal:2010ig}, we have enough constraints to perform a global fit to these data within the SHM framework. In Appendix~\ref{apx:RHIC62}, we perform a detailed analysis using the yield of $\pi^\pm, K^\pm, K_S^0, p^\pm, \phi, \Lambda, \overline{\Lambda},\Xi^\pm$ and $\Omega^\pm$ as a function of centrality. We use the number of participating nucleons $N_{\mathrm{npart}}$ as a measure of centrality characterizing the impact factor of the collision, see Ref.~\citep{Miller:2007ri} for centrality measure overview. We overcome the problem of the data reported in incompatible centrality bins by interpolating each hadron species with a power law in the form 
\begin{equation}
\label{eq:interpolation}
\frac{dN_i \left( N_{\mathrm{npart}} \right) }{dy} = a_i \, N_{\mathrm{npart}} ^{b_i} + c_i,
\end{equation}
where the parameters $a_i,b_i,c_i$ are fitted to best describe the centrality dependence of particle $i$. This very precise empirical description enables us to perform an SHM fit to all the particle species for arbitrary centrality. 

It is noteworthy that the publication of experimental multistrange particle data~\citep{Aggarwal:2010ig} included a semi-equilibrium fit to the data, constrained however by $\gamma_s<1,\gamma_q=1$. Our analysis with the semi-equilibrium model (fixed $\gamma_q=1$) of the same data set and unconstrained parameter $\gamma_s$ values leads to compatible results only for peripheral collisions, where $\gamma_s$ converges to values below unity. For central, head-on collisions, the measured strange hadron yields correspond to $\gamma_s>1$, an overpopulated strangeness phase space. The hadronization temperature assuming semi-equilibrium approach converges to $T^{\mathrm{semi}}\simeq 160\,\mathrm{MeV}$. 

When we release the light quark phase space occupancy $\gamma_q$ and fit the data within the non-equilibrium model, we find that the light quark phase space is also overpopulated. This is described by the centrality independent value $\gamma_q\simeq 1.6$. The centrality independence is expected from the analysis of the $\Xi/\phi$ ratio shown in the previous Section~\ref{sec:ratios}, assuming that the hadronization temperature $T$ is constant. In full non-equilibrium model, strangeness is even further from the absolute equilibrium value reaching $\gamma_s=2.2$ for the most central collisions.

Another very important aspect of the SHM fit shown is the opportunity we have to examine the physical properties of the bulk at hadronization. We expect the same hadronization conditions of the bulk irrespective of the relativistic heavy-ion collision origin and experimental setup leading to the QGP fireball, e.g., determining the volume and the initial collision energy.
The physical properties of the bulk, the energy density $\varepsilon$, the pressure $P$, and the entropy density $\sigma$ are of particular interest in the the study of QGP hadronization. All three vary significantly as a function of centrality in the semi-equilibrium model, whereas in the chemical non-equilibrium approach these bulk properties are virtually constant for all centralities. This result strongly supports the non-equilibrium SHM variant within the context of the universal QGP hadronization conditions hypothesis. We cannot exclude fitting the RHIC 62 GeV data with the semi-equilibrium model based on the fit quality, as the large data uncertainties are tolerant enough to accommodate $\gamma_q=1$. However, the non-equilibrium is clearly preferred on physics grounds; it leads to compatible physical properties of different size fireballs at break-up.

A critical hadronization pressure $P=82\,\mathrm{MeV/fm}^3$ has been proposed for heavy-ion collisions at SPS energies~\citep{Rafelski:2009jr}. We find the same hadronization pressure for all centralities within the non-equilibrium SHM fit at a CM energy almost 4 times higher than the top SPS energy and for fireballs spanning  more than an order of magnitude in reaction volume. This suggests that the intensive bulk properties we evaluated for this system: critical hadronization pressure $P=82\,\mathrm{MeV/fm}^3$, energy density $\varepsilon = 0.50\,\mathrm{GeV/fm}^3$, and entropy density $\sigma = 3.3\,\mathrm{fm}^{-3}$, are in fact nearly universal hadronization conditions for QGP created in heavy-ion collisions.

\section{Hadronization at LHC}
\label{sec:HadronizationAtLHC}
\subsection{Does SHM describe particle production at LHC?}

\begin{figure}[t!]
\centering
\includegraphics[width=0.6\columnwidth]{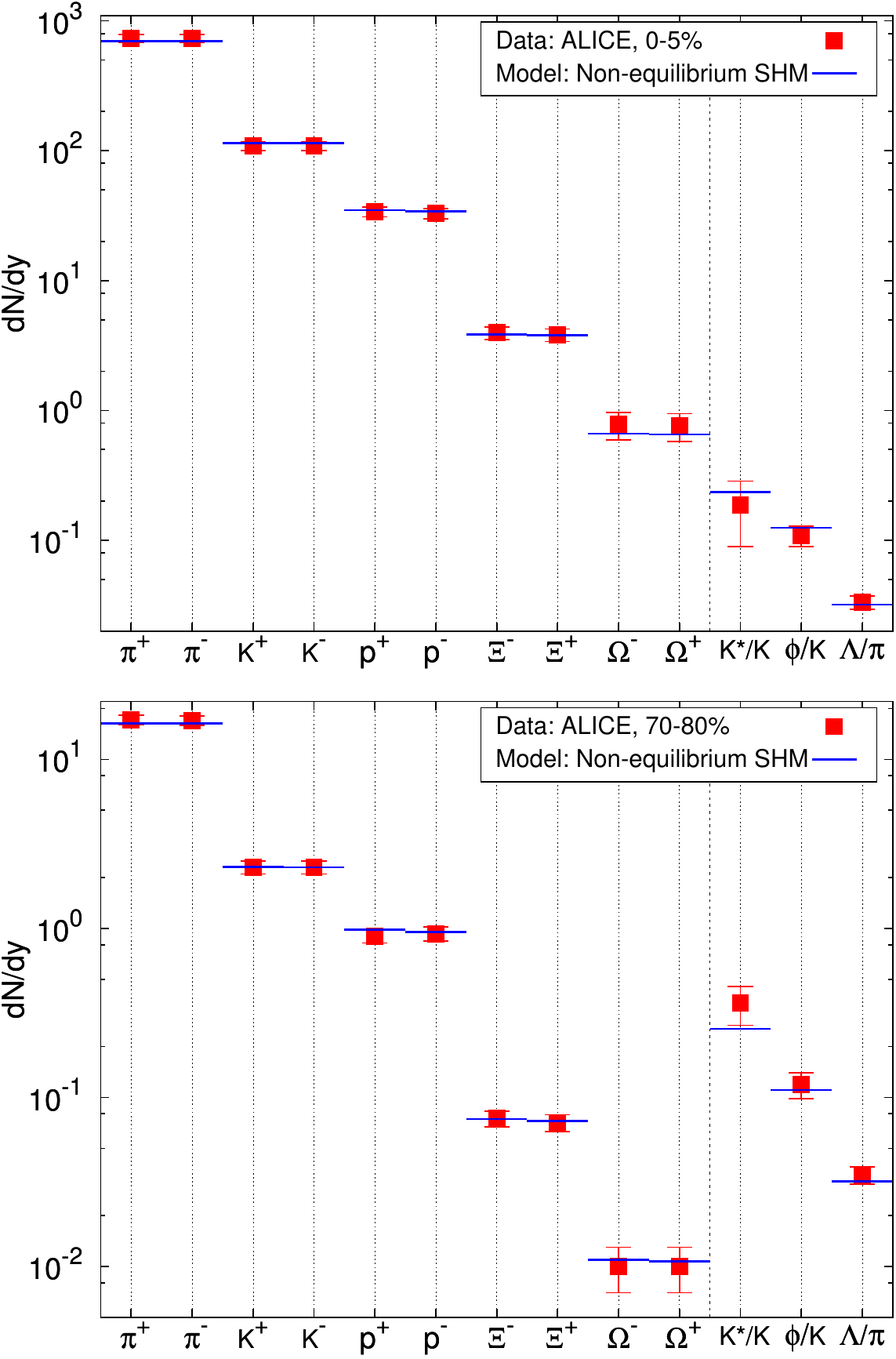}
\caption[SHM fit to data from Pb--Pb collisions at $\sqrt{s_{NN}}=2.76\,\mathrm{TeV}$]{\label{fig:FitLHC0020}Non-equilibrium SHM fit to experimental data from 0--5\% and 70--80\% centrality Pb--Pb collisions at $\sqrt{s_{NN}}=2.76\,\mathrm{TeV}$ measured by the ALICE experiment. Note that the model describes particle yields from $dN/dy\sim 1000$ to $dN/dy\sim 0.01$.}
\end{figure}

A na\"{i}ve semi-equilibrium SHM fit has been presented together with the preliminary data from 0--20\% central Pb--Pb collisions at $\sqrt{s_{NN}}=2.76\,\mathrm{TeV}$ measured by the ALICE experiment~\citep{Ivanov:2013haa}. The ALICE experiment has higher precision tracking than RHIC, which leads to smaller experimental data uncertainties. The simple fit presented in~\citep{Ivanov:2013haa} yields inconsistencies when fitting different hadron species, creating the impression in the community that SHM does not work at LHC. This impression was caused by the constrained $\gamma_s<1, \gamma_q = 1$ semi-equilibrium model, which does not agree (as it does at RHIC) with all particle production data from heavy-ion collision at LHC energy. The largest reported inconsistency is the ratio of protons to pions $p/\pi$. Recalling the method from Section~\ref{sec:ratios}, we see that according to the quark content, this ratio has to be proportional to $p/\pi\propto \gamma_q$. In Appendix~\ref{apx:UniversalHadronization}, we successfully fit the same data as in~\citep{Ivanov:2013haa}, applying the non-equilibrium SHM that allows $\gamma_q$ to vary, permitting also $\gamma_s>1$. The correct experimental value of $p/\pi$ ratio is a natural outcome of the fit with a value of $\gamma_q \simeq 1.6$, as predicted in~\citep{Rafelski:2010cw}. Similarly to RHIC results confirming the need for $\gamma_s > 1$, at this point the LHC results require both $\gamma_q > 1$ and $\gamma_s > 1$ . Moreover, as Figure~\ref{fig:FitLHC0020} shows, the non-equilibrium model accurately describes all hadron production from central to peripheral collisions that is over 5 orders of magnitude in their yields, where the most variable parameters as a function of centrality are the normalization $dV/dy$ and strangeness phase space occupancy $\gamma_s$.

We complement the preliminary data from the central Pb--Pb collisions with centrality dependent yields of $\pi^\pm, K^\pm$ and $p^\pm$ from~\citep{Abelev:2013vea} and ratios $\phi/K$, $\Lambda/\pi$ and $K^{0*}/K^-$ from~\citep{Singha:2012qv}. We perform a centrality dependent analysis  with a focus on the hadronization conditions at an energy 45 times higher than RHIC 62, the previous system we analyzed. In order to unify the centrality binning, we use four centrality bins in which $K^{0*}/K^-$ is reported. We match the other data by taking the average of two neighboring points. 

This centrality analysis reveals that the main difference from RHIC 62 is a 4 times larger volume of the hadronizing fireball, see Figure~\ref{fig:VolTemp} below. Hadronization temperature $T$ is found to be a few MeV lower for most central collisions; baryochemical potential is difficult to quantify, but is of the order of 1 MeV. The strangeness phase space, while showing a large overpopulation, is 20\% lower than expected based on RHIC 200 results analysis and extrapolation~\citep{Rafelski:2010cw}. Those results suggest a long lifespan of the fireball created at LHC, allowing strangeness to annihilate to a lower final abundance. The intensive properties at freeze-out are very similar to those at RHIC: pressure $P=82\pm 8\,\mathrm{MeV/fm}^3$, energy density $\varepsilon=0.50\pm 0.05\,\mathrm{GeV}$ and entropy density $\sigma = 3.35 \pm 0.30\,\mathrm{fm}^{-3}$. This result confirms the existence of universal hadronization conditions of the QGP fireball, which we introduced in the previous Section~\ref{sec:StrangenessAtRHIC62} and Appendix~\ref{apx:RHIC62}.

\begin{figure}[t]
\centering
\includegraphics[width=0.9\columnwidth]{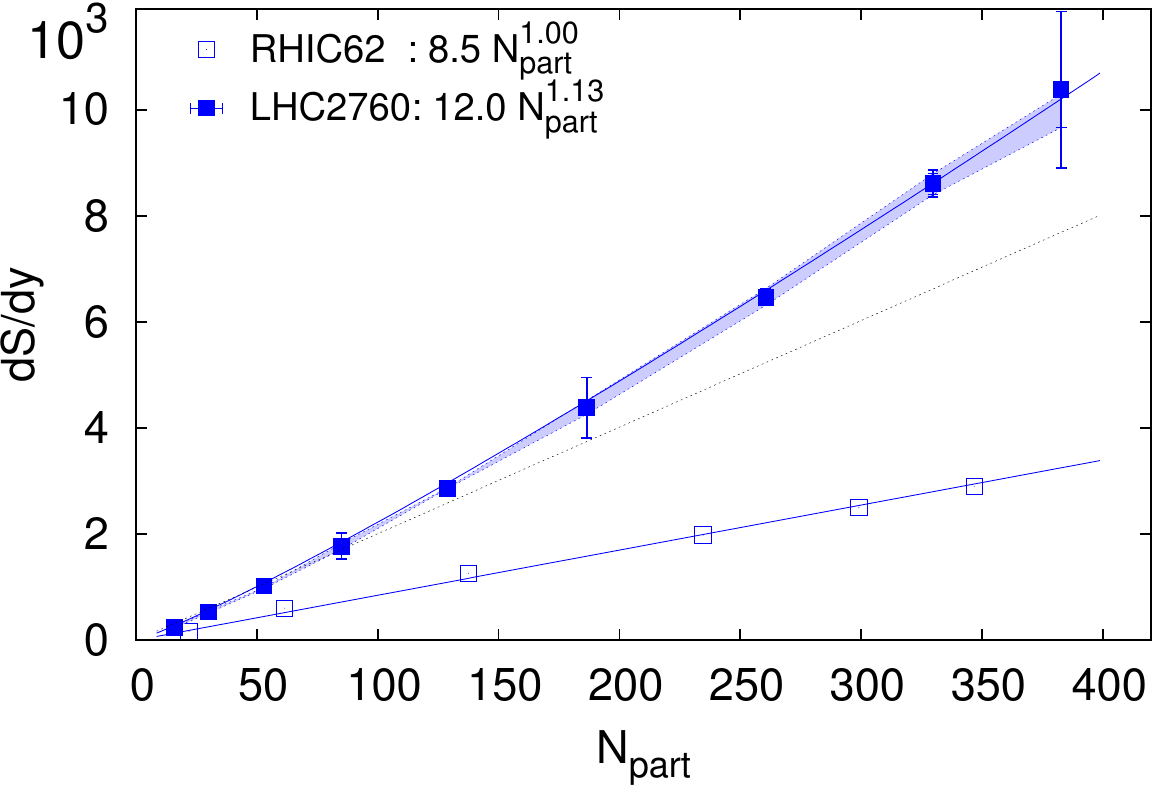}
\caption[The entropy yield at LHC and RHIC as a function of centrality]{\label{fig:entropy}The entropy yield at LHC and RHIC as a function of centrality. The black dotted line shows a linear growth, which is to be compared with the top LHC line.}
\end{figure}

In Figure~\ref{fig:entropy} we show an unexpected behavior of the entropy content of the fireball. The entropy yield of the QGP fireball at LHC increases more steeply with the number of participants than the linear behavior seen at RHIC 62, which implies an additional entropy production mechanism proportional to centrality of the collision. At the present time, we suspect charm re-annihilation in the QGP and charm decays after hadronization to be the extra omitted entropy production mechanism, a phenomenon we plan to explore as an application of SHARE with CHARM program we developed, see Appendix~\ref{apx:ShareManual}.

\subsection{Centrality independence and chemical equilibrium of light quarks}
In Appendix~\ref{apx:Alice2760}, we present a detailed comprehensive centrality dependent analysis of hadron production in Pb--Pb collisions from LHC. We complement the data set used in the first analysis presented in Appendix~\ref{apx:UniversalHadronization} with multistrange baryons $\Xi$ and $\Omega$ and unify centrality binning using the interpolation method developed for RHIC data analysis discussed earlier in Section~\ref{sec:StrangenessAtRHIC62}. This enables us to perform a detailed non-equilibrium SHM analysis in 9 centrality bins and to discuss and reject the alternative models.

\begin{figure}[t]
\centering
\includegraphics[scale=0.85]{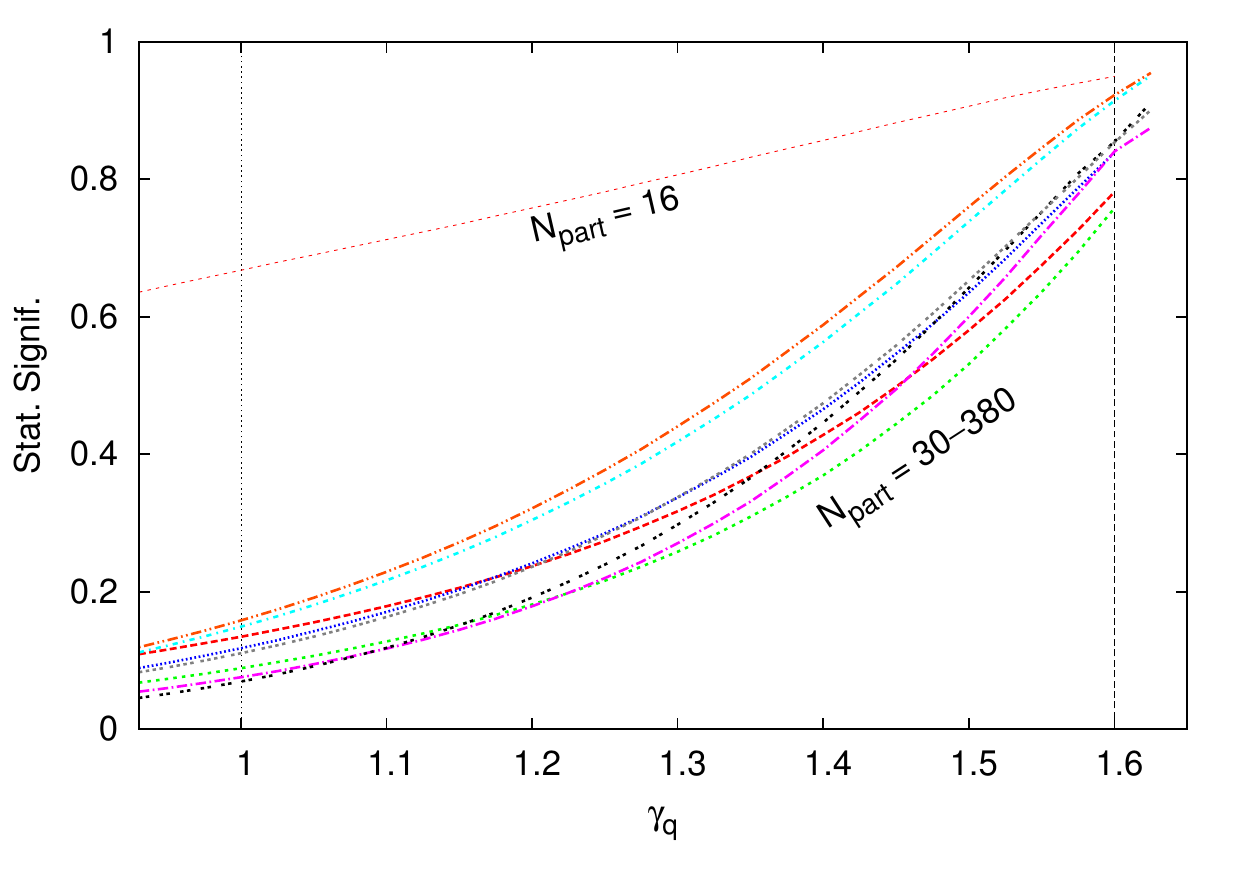}
\caption[Statistical significance profile of $\gamma_q$]{\label{fig:gamq-profile}Statistical significance profile of the best fit to data from Pb--Pb collisions at $\sqrt{s_{NN}}=2.76\,\mathrm{TeV}$ as a function of $\gamma_q$ for all studied centrality bins.}
\end{figure}

We thoroughly compare fits to the data with three SHM equilibrium variants; equilibrium ($\gamma_s=\gamma_q=1$), semi-equilibrium ($\gamma_s\neq 1, \gamma_q=1$), and non-equilibrium ($\gamma_s\neq 1, \gamma_q\neq 1$). Recall that the equilibrium model has been proven inapplicable already at RHIC energies. The more precise LHC data clearly prefers non-equilibrium with $\gamma_q\simeq 1.6$ , as we expect based on the constant $\Xi/\phi$ ratio, which keeps the same value at RHIC and LHC, see Figure~\ref{fig:XiPhi-ratio}. We show the necessity of $\gamma_q > 1$ by evaluating the statistical significance profile as a function of $\gamma_q$ shown in Figure~\ref{fig:gamq-profile}. This result clearly shows; first, that $\gamma_q=1$ has no special meaning for the model and has the lowest confidence level; and second, that the highest confidence level corresponds to the value of $\gamma_q$ near its critical value of Bose-Einstein condensation of $\pi^0$ defined by 
\begin{equation}
\label{eq:gamqcrit}
\gamma_q^{crit} = \exp\left(\frac{m_{\pi^0}}{2T}\right).
\end{equation}
The only qualitatively different behavior is the most peripheral bin we analyze, 70--80\% ($N_{\mathrm{npart}}\simeq 16$), for which the $\chi^2$ of the fit is more tolerant to $\gamma_q < 1.6$. One can infer that in peripheral collisions, for example, surface contribution could play an important role~\citep{Aichelin:2008mi}.

With prescribed $\gamma_q=1$, the total $\chi^2$ of the fit increases by a factor of 4 as compared to $\gamma_q \sim 1.63$, confirming the need for $\gamma_q$ as a free model parameter. Although one can argue that adding a free parameter to a model automatically improves fit to the data, it is noteworthy that releasing only $\gamma_s$ does not improve the fit decisively, as does releasing and fitting $\gamma_q$ and $\gamma_s$. On the other hand, we have already presented arguments based on the study of particle ratios that $\gamma_s\neq 1$, see Section~\ref{sec:ratios} and Appendix~\ref{apx:MultistrangeRatios}. We thus conclude that $\gamma_q\neq 1$ is required as well.

We further exclude models in which the hadron abundance evolve strongly after hadronization (for example the annihilation process $p+\bar{p} \to $ pions) on the grounds of wrong centrality dependence. We find that equilibrium SHM complemented by such `afterburner' does not describe the data adequately. The main challenge of this effort has been the measured ratio of $p/\pi=0.046$ compared to the factor of $1.5-1.9$ larger value predicted by the equilibrium SHM~\citep{Abelev:2012wca}, which could be reduced via the $p\bar{p}$ annihilation. Although the posthadronization interactions can explain this isolated data point in central collisions, they have two huge challenges to overcome before becoming a viable model. First, the annihilation interactions are highly centrality dependent; the ratio increases to $p/\pi=0.058$ already in 20--30\% ($N_{\mathrm{npart}}=186$) centrality bin~\citep{Karpenko:2012yf}, which is in contradiction by 3 standard deviations with the experimentally measured centrality independent $p/\pi$ ratio. Second, while solving the `proton anomaly' with baryon-antibaryon annihilation, the posthadronization interactions cause a new anomaly in the equilibrium model value of $\Xi$ yield, which is depleted by these interactions disagreeing further with the experimental data.

\begin{figure}[t]
\centering
\includegraphics[width=\columnwidth]{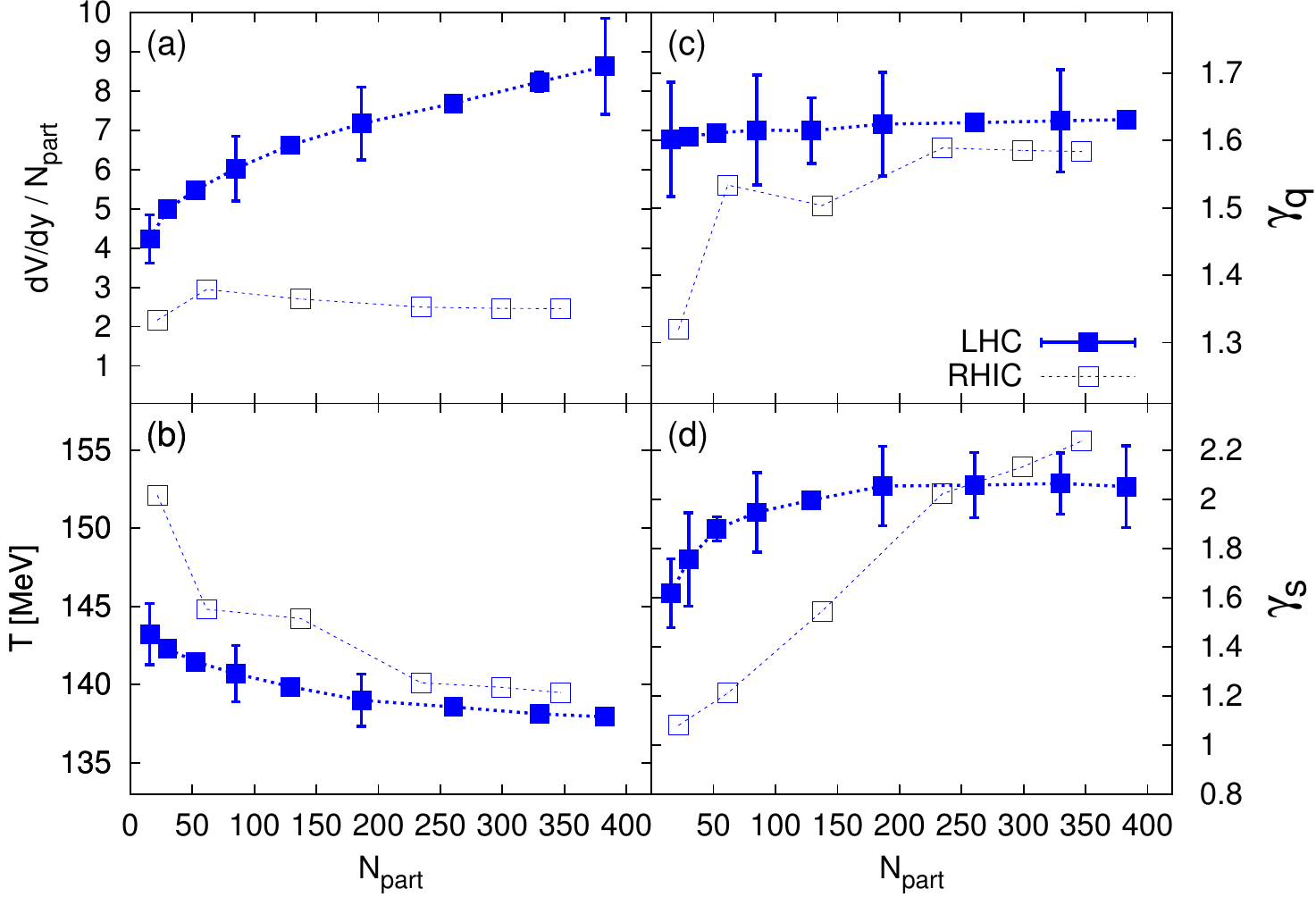}
\caption[Centrality dependent SHM parameters fitted to data from LHC ocmpared to RHIC]{\label{fig:VolTemp}SHM parameters obtained by fitting the data from Pb--Pb collisions at $\sqrt{s_{NN}}=2.76\,\mathrm{TeV}$ from LHC (Appendix~\ref{apx:Alice2760}) compared to RHIC 62 (Appendix~\ref{apx:RHIC62}). Volume normalized to number of participants in panel (a), freeze-out temperature $T$ in panel (b), light quark phase space occupancy $\gamma_q$ in panel (c) and strangeness phase space occupancy $\gamma_s$ in panel (d).}
\end{figure}

Having excluded other model approaches, we return to elaborate on the results of SHM non-equilibrium analysis. In Figure~\ref{fig:VolTemp}, we show a comparison of centrality dependent non-equilibrium SHM parameters for LHC and RHIC62. Panel (a) shows the volume per unit rapidity normalized to the number of participating nucleons $(dV/dy)/N_{\mathrm{npart}}$; we note the rise with centrality. The freeze-out temperature $T$ (panel (b) of Figure~\ref{fig:VolTemp})  is very close in the two environments, RHIC and LHC. We observe a temperature drop of a few MeV, which we associate with deeper supercooling of a more dynamically expanding fireball at the LHC energy. However, it is possible that systematic differences in detector properties (ALICE vs. STAR) are the origin of this difference, which is within the fit error. The light quark phase space occupancy $\gamma_q$ (panel (c) of Figure~\ref{fig:VolTemp}) saturates earlier, i.e., for more peripheral systems at LHC. The strangeness reaches a steady level of equilibration for relatively small systems ($N_{\mathrm{npart}}\gtrsim 50$), as seen in the value of the strangeness phase space occupancy $\gamma_s\to 2$.

\begin{figure}[t]
\centering
\includegraphics[width=0.9\textwidth]{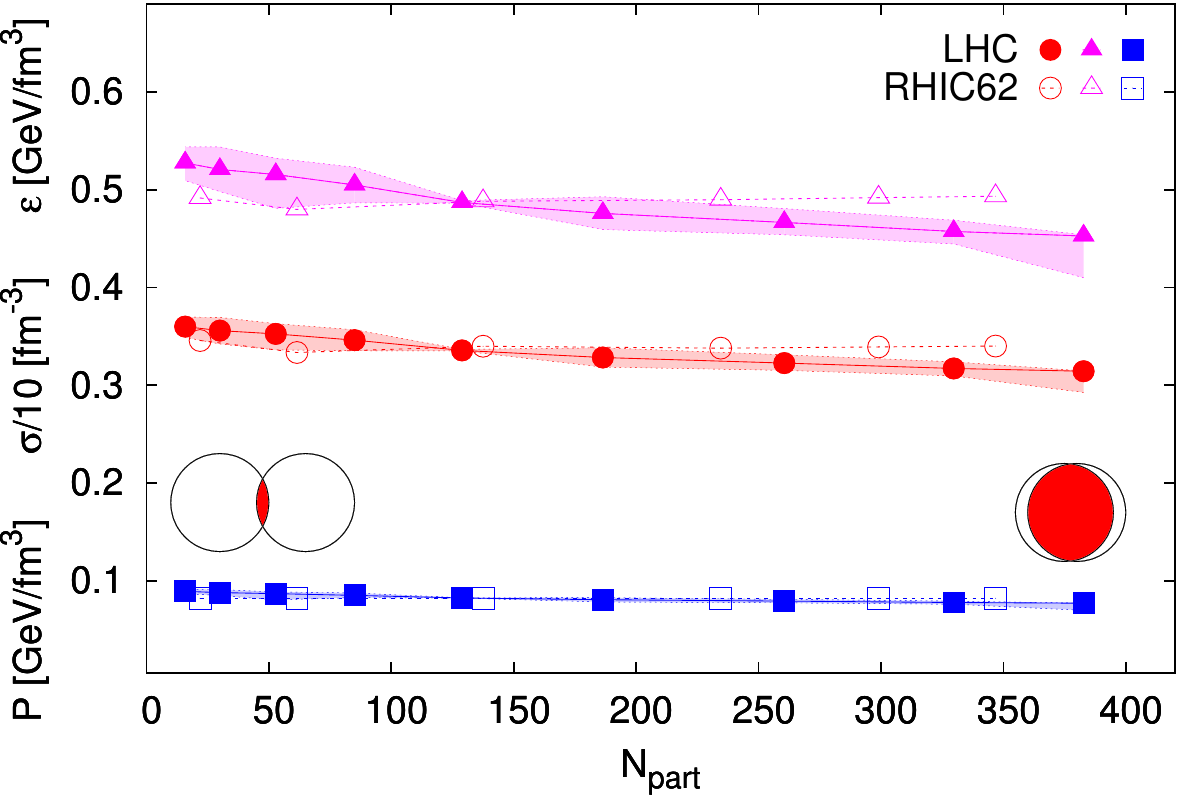}
\caption[Physical properties of the hadronizing QGP fireball at RHIC62 and LHC]{\label{fig:BulkProperties}Physical bulk properties of the hadronizing QGP fireball from RHIC62 and LHC, critical pressure $P$, energy density $\varepsilon$, and entropy density $\sigma$, as a function of centrality.}
\end{figure}

As data in Figure~\ref{fig:XiPhi-ratio} show, hadron ratios proportional to $\gamma_s$ are also almost constant at LHC compared to lower energy experiments. This behavior is consistent with the resulting chemical parameters of the SHM fit; in particular the strangeness phase space saturates at $\gamma_s\simeq 2$ already for rather peripheral collision, $N_{\mathrm{npart}}\geq 50$, see Figure~\ref{fig:VolTemp}(d). This shows that the high initial energy density fireball (increasing greatly the final volume $V$) lives sufficiently long for strangeness to establish a stable chemical equilibrium in the QGP. 
This QGP equilibrium of strangeness implies chemical non-equilibrium of produced hadrons~\citep{Letessier:2002gp}.

In Appendix~\ref{apx:Alice2760}, we further evaluate the physical bulk properties of the QGP fireball at LHC and compare them to our previous results from RHIC. New additional data and finer centrality binning are compatible with results also shown in Appendix~\ref{apx:UniversalHadronization} further confirming the conclusion that quark-gluon plasma created in heavy-ion collisions at TeV energies hadronizes at nearly the same universal hadronization conditions as at RHIC (and therefore also SPS), see Figure~\ref{fig:BulkProperties}.

\subsection{Fit stability, omitted resonances and strange quark mass at hadronization}

In Appendix~\ref{apx:StrangeSQM}, we study new ALICE data and the physical stability of our fit, and therefore the physics reliability, of the chemical non-equilibrium SHM fits, that we presented above and in Appendices~\ref{apx:UniversalHadronization} and~\ref{apx:Alice2760}. We first confirm that the newly published final data of $K_S^0, \Lambda, \Xi$ and $\Omega$ agree practically exactly with our presented interpolations and estimates. Furthermore, we perform a fit with finite hadron resonance widths, which in general can cause a shift in computed particle yields due to resonances being created more abundantly below their central mass. We find the calculation with and without finite widths to be virtually the same, except for an overall normalization change which is compatible within the fit errors of our earlier work.

Detailed inspection of the individual fitted data points reveals that the only systematic difference between the model and the experimental yield is for $\Lambda$ (e.g., $\Lambda = 17\pm2$ in the 10--20\% centrality bin) which is systematically underpredicted: the model value is a little over 1 standard deviation below the experimental point for all centralities. This may mean that the deviation is due to a systematic centrality independent omission in our model.  We explore the Particle Data Book~\citep{Beringer:1900zz} seeking a 2-star (**) resonance that, upon inclusion in the SHARE particle list, could serve as a significant source of $\Lambda$. We identify $\Sigma(1560)$ as a potential candidate. The charged states $\Sigma(1560)^\pm$ have been observed with $6\sigma$ signal; however, the neutral state $\Sigma(1560)^0$ has not been confirmed by an independent experiment. This resonance has only one decay channel $\Sigma(1560) \xrightarrow{\;\:100\%\;\;} \Lambda\pi$; it is light enough and likely has spin \sfrac{3}{2}, so it could be produced abundantly enough to provide the missing source of $\Lambda$. After including the $\Sigma(1560)$ and its decay in the particle list, the subsequent SHM fit converges to the same SHM parameter values. The only change is a significantly lower $\chi^2$, because the $\Lambda$ yield is now fitted within \sfrac{1}{2} standard deviation below the experimental point, and thus contributes significantly less towards the total $\chi^2$. 

After establishing the fit compatibility with the new final data and fit stability against finite resonance width, and against an addition to the hadron mass spectrum, we turn to address strangeness conservation during a sudden hadronization process. We evaluate the observed strangeness in hadrons and match it to the strangeness in QGP. We offer two possible extreme scenarios:
\begin{enumerate}
\item Strangeness is chemically equilibrated in the QGP created in central collisions. This, however, requires the strange quark to acquire an effective mass of $m_s=299\,\mathrm{MeV}/c^2$.
\item Strange quark has its PDG mass $m_s = 140\,\mathrm{MeV}/c^2$ (at the relevant scale $\mu\simeq 2\pi T=0.9\,\mathrm{GeV}$) and strangeness phase space in the QGP is undersaturated with $\gamma_s^{\small QGP}=0.77$.
\end{enumerate}
Further analysis may bring these two scenarios together by, e.g., incorporation of QCD many body interactions and/or longitudinal expansion strangeness dilution.

\section{Charm production and hadronization}

\begin{figure}[H]
\centering
\includegraphics[angle = 90, height=\textheight]{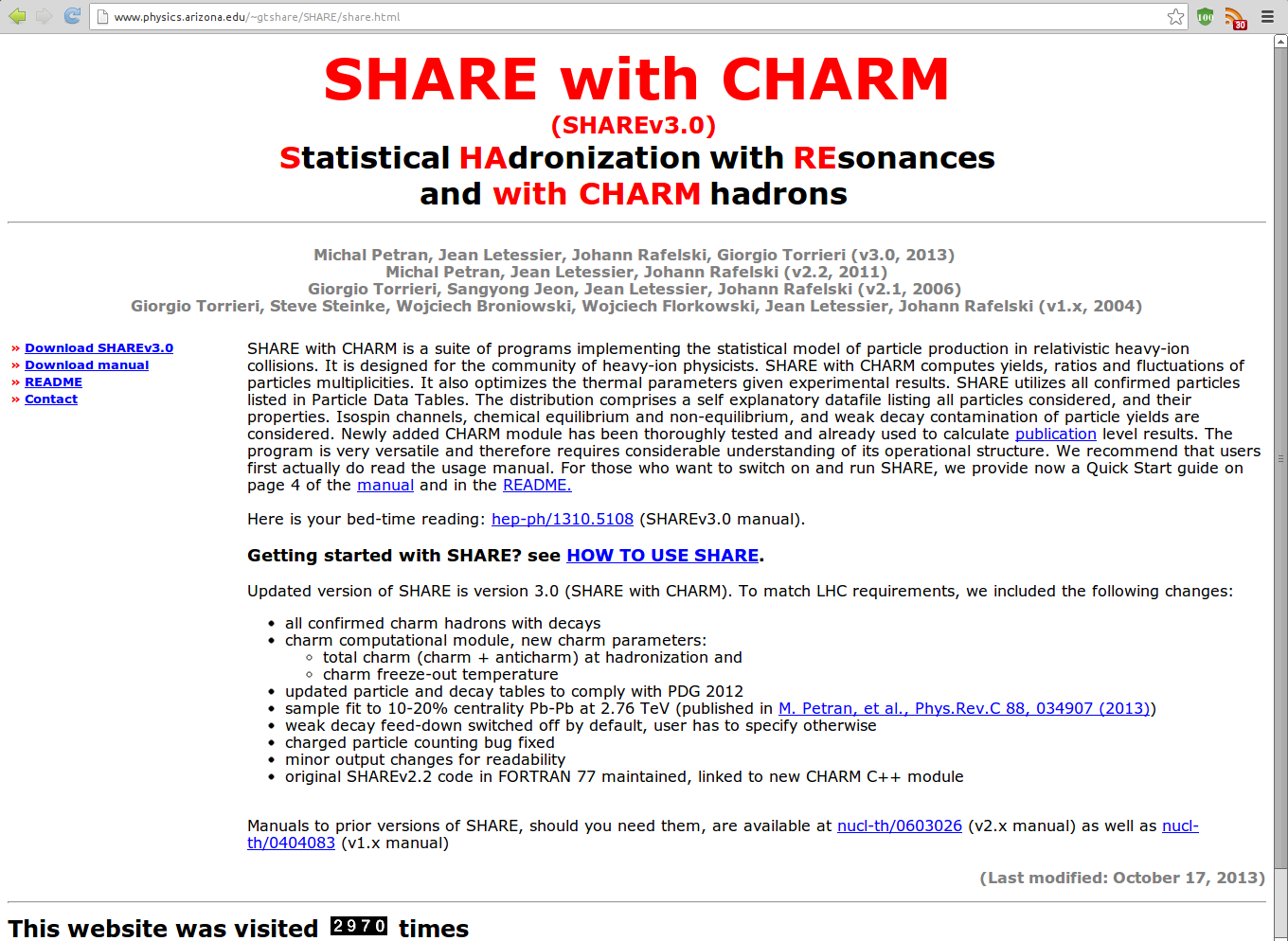}
\caption[Screenshot of SHARE with CHARM webpage]{\label{fig:SHAREwebpage}Screenshot of SHARE with CHARM webpage.}
\end{figure}

With a large yield of charm expected in heavy-ion collisions at LHC, hadronization of charm becomes a new necessary feature to be accounted for in the SHM. For this purpose, we developed and present a tool in Appendix~\ref{apx:ShareManual}: SHARE with CHARM. It builds on the success of its predecessors, SHAREv1~\citep{Torrieri:2004eb} and SHAREv2~\citep{Torrieri:2006xi}. SHARE was developed at the Department of Physics of the University of Arizona in collaboration with Cracow and Montreal. The first release, SHAREv1~\citep{Torrieri:2004eb}, introduced hadron yields and yield calculations for the light ($u,d,s$) quarks. SHAREv2~\citep{Torrieri:2006xi} added calculation of event-by-event fluctuations and has been widely used within the community as a tool for many publication results. SHARE with CHARM (SHAREv3) described in Appendix~\ref{apx:ShareManual} now enhances the powerful implementation of SHM with charm capabilities and provides an effective tool for SHM studies of particle production in the LHC era. The program is publicly available at \url{http://www.physics.arizona.edu/~gtshare/SHARE/share.html} for download together with operation instructions. For the SHARE with CHARM website screenshot, see Figure~\ref{fig:SHAREwebpage}.

SHARE with CHARM is designed to address the following questions:
\begin{itemize}
\setlength{\parskip}{0mm}
\setlength{\itemsep}{0mm}
\item What number of charm quarks is present at hadronization?
\item Is charm hadron production in heavy-ion collisions described by SHM?
\item Does charm freeze-out at the same temperature as other hadrons?
\item Are the physical properties of the hadronizing fireball affected by a non-negligible amount of charm present at hadronization?
\end{itemize} 

SHARE with CHARM will help to answer all of the above questions once charm hadron yield data from heavy-ion collisions at LHC become available. Currently, we have implemented in the program all 3-star (***) and 4-star (****) hadrons and resonances, we also included states that were not yet confirmed experimentally, but must exist for symmetry reasons of the quark model, such as $\Omega_{ccc}^{++}$, a bound state of three charm quarks $(ccc)$. Creating a charm hadron decay table is an even more challenging task. Considering the high number (a few hundred in some cases) of decay channels with comparable branching ratios, we have to include all decay channels with a relative branching ratio over $10^{-4}$, sometimes $10^{-5}$, when no dominant channel is present. Oftentimes, due to their high mass, charm hadrons decay into four, five or more daughter particles. For many of the decay final multiparticle states, an isospin symmetric decay product state may not have been observed, because some of them are very challenging to measure experimentally. We add these into the charm decay tree, as they are the most likely candidates of unobserved decay channels allowing the sum of branching ratios to add to one. 

As an example of this procedure, consider one of the dominant and measured decay channels of $\Lambda_c^+$:
\begin{equation}
\label{eq:LambdaCdecay}
\Lambda \rightarrow p\overline{K^0}\pi^0 \qquad \mathrm{BR}=(3.3\pm1.0)\%.
\end{equation}
We add in the decay table a channel symmetric to this one with the same branching ratio
\begin{equation}
\label{eq:LambdaCdecay2}
\Lambda \rightarrow n\overline{K^0}\pi^+ \qquad \mathrm{BR}=(3.3\pm1.0)\%,
\end{equation}
even though this channel was not experimentally observed. The decay channels added by hand and the relatively large uncertainties of the branching ratios introduce an intrinsic uncertainty into our CHARM module, which is therefore subject to future refinement as more precise data inputs become available in the future PDG releases and as additional theoretical constraints are recognized.

The new input files and the external CHARM program module are responsible for charm hadron yields calculation based on the thermal parameters set by SHARE. The two parameters specific to charm are: the total number of charm and anticharm quarks $N_{c\bar{c}}=c+\bar{c}$; and the ratio of charm hadronization temperature to the freeze-out temperature of light hadrons $T_{\mathrm{charm}}/T$. The CHARM module then calculates the  charm feed-down into hadron yields, solves for $\gamma_c$ and feeds all information back into SHARE for subsequent calculations.

In Appendix~\ref{apx:CharmSQM}, we present the immediate effects of charm hadronization for a range of produced charm $N_{c\bar{c}}$ based on the estimates described above in Section~\ref{sec:charm-estimates}. In central Au--Au collisions at $\sqrt{s_{NN}}=200\,\mathrm{GeV}$ at RHIC, there is relatively little charm, $N_{c\bar{c}}=8.6$. We find the effect on SHM fit with charm is small at RHIC in terms of modification of both final fit parameters and physical bulk properties of the hadronizing fireball. The effect is well within the experimental errors of the data available for this system.

For LHC, we base our study on the two charm production estimates mentioned in Section~\ref{sec:charm-estimates}; that is the charm production cross section scaling indicating $N_{c\bar{c}} = 246\pm 154$ (large error due to $pp$ charm cross section uncertainty), and the incomplete $D^0$ meson $p_\perp$-spectrum consistent with $N_{c\bar{c}} = 10 - 60$. During this study, we fit the non-charm hadron data from LHC used in our previous work (see Appendix~\ref{apx:Alice2760}), while prescribing a fixed amount of charm in the range of $N_{c\bar{c}} \in (0,400)$. This way we cover the range of charm production up to the forthcoming energy upgrade of LHC. In this preliminary exploration, we kept the charm hadronization temperature equal to that of other hadrons.

\begin{figure}[t]
\centering
\includegraphics[scale=1.]{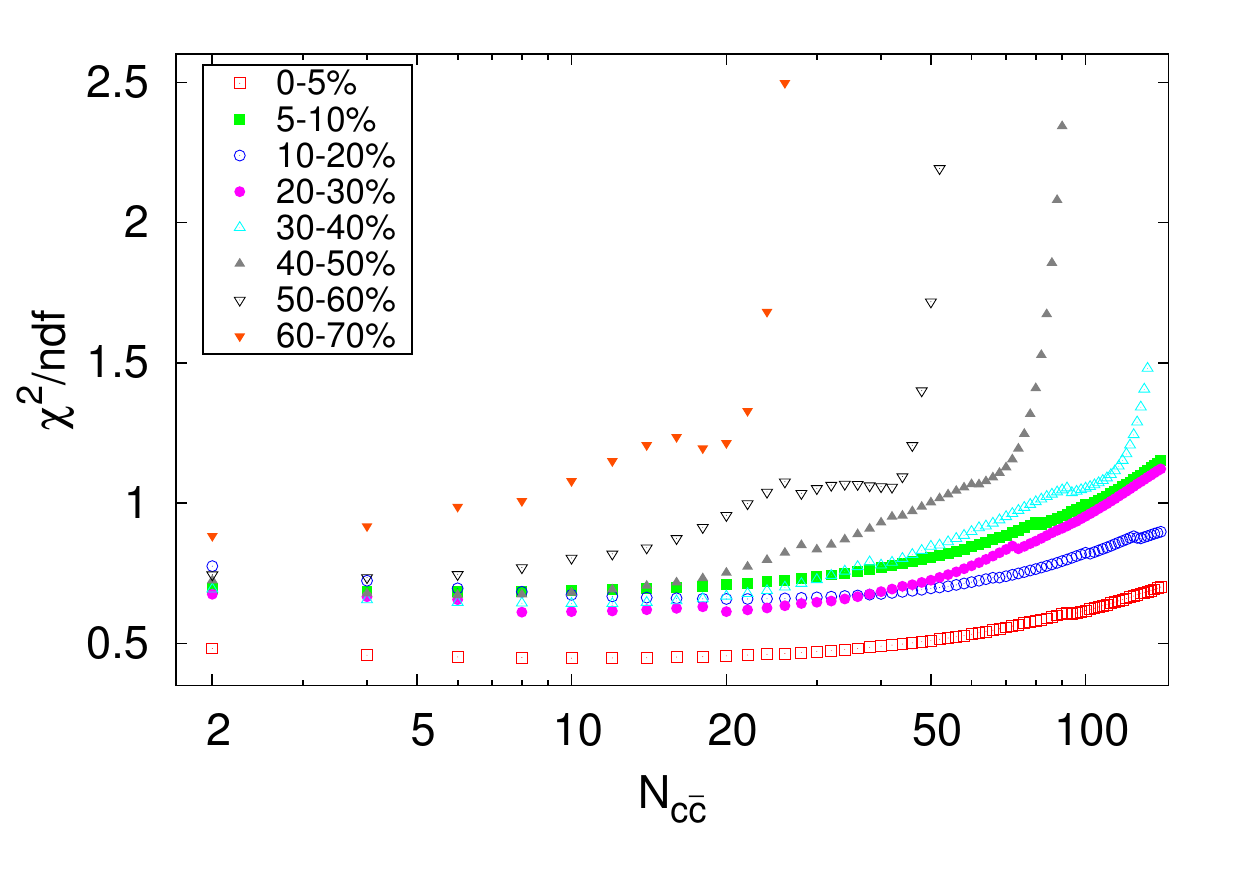}
\caption[$\chi^2$ profile of fits to LHC data for different charm amount $N_{c\bar{c}}$]{\label{fig:Ncbc-profile}$\chi^2$ profile of fits to non-charm data from Pb--Pb collisions at $\sqrt{s_{NN}}=2.76\,\mathrm{TeV}$ from LHC (analyzed in Appendix~\ref{apx:Alice2760}) as a function of total charm present at hadronization $N_{c\bar{c}}=c+\bar{c}$.}
\end{figure}

The charm hadron feed-down represents, in general, a different pattern of hadron species production. We test if the non-charm data from LHC, which we collected for Appendix~\ref{apx:Alice2760}, constrain the charm abundance through the charm decay pattern. For each centrality we calculated a $\chi^2$ profile in $N_{c\bar{c}}$. As a result, we obtain a range of compatible $N_{c\bar{c}}$ for each centrality that does not contradict the non-charm hadron data, see Figure~\ref{fig:Ncbc-profile}. For instance, in the 50--60\% centrality bin, $N_{c\bar{c}} < 50$ is compatible with the data; whereas the more peripheral bin 60--70\% suggests a tighter limit, $N_{c\bar{c}} < 25$.  However, no clear minima of the $\chi^2$ function in the compatible $N_{c\bar{c}}$ ranges are observed. We conclude that in absence of a directly measured charm hadron yield, charm decay hadron yields impose only a weak constraint on the total amount of charm, and that future data are likely to be compatible with our current SHM results.

\begin{figure}[t]
\centering
\includegraphics[width=0.9\columnwidth]{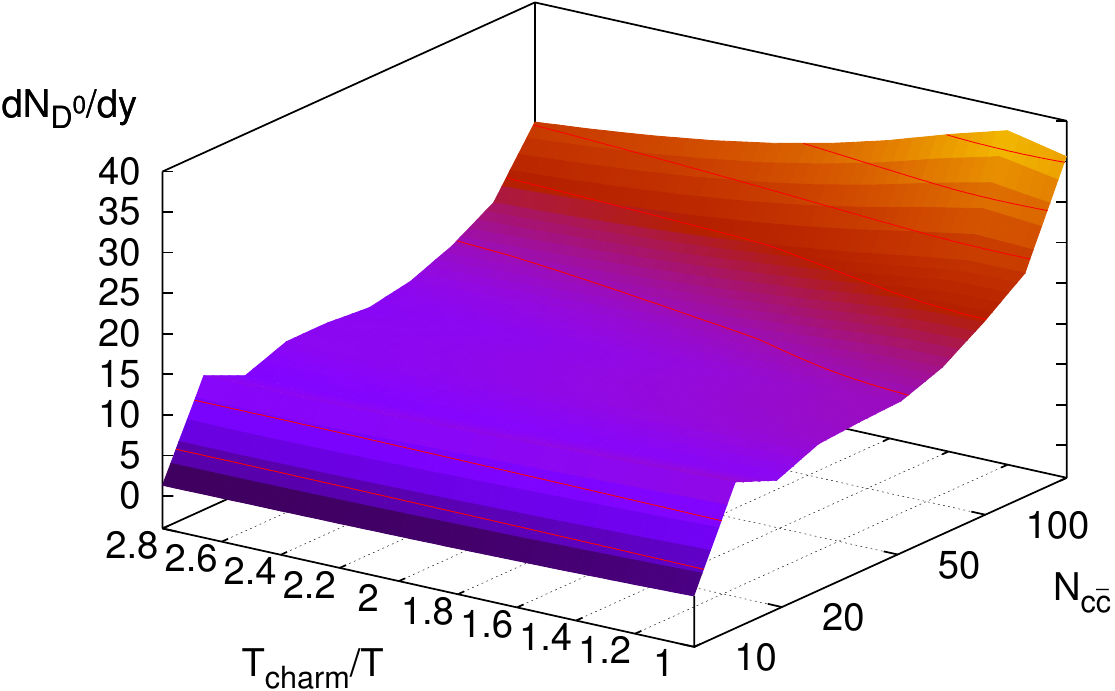}
\caption[$D^0$ yield as a function of $N_{c\bar{c}}$ and $T_{\mathrm{charm}}/T$]{\label{fig:D03d}$D^0$ yield as a function of charm abundance at hadronization $N_{c\bar{c}}$, and ratio of charm hadronization temperature to that of other flavors $T_{\mathrm{charm}}/T$. Solid lines on the surface depict constant yield of $D^0$.}
\end{figure}

We observe that charm hadron yields scale with $N_{c\bar{c}}$. While $D^0(c\bar{u})$ meson scales almost linearly, the yield of, for instance, $\Xi^{++}_{cc}(ccu)$ is proportional to $N_{c\bar{c}}^2$. Recalling that a yield is proportional to one power of $\gamma_f$ for every constituent (anti)quark of flavor $f$, we can attribute this proportionality to $\gamma_c$ assuming large values, and thus playing a dominant role in charm hadron yield calculation. The yields of single charm hadrons are well described within our framework irrespective of the $T_{\mathrm{charm}}$ value. This effect is further discussed below, see also Figure~\ref{fig:D03d}. A more complete test of SHM with charm comes when several multicharm hadrons will have been measured.

Further, we reach the same conclusion as~\citep{Kuznetsova:2006bh}, that charm decays can be a significant source of multistrange particles. This is primarily caused by the enhanced strangeness present in the fireball at hadronization making it very likely for a charm quark to bind to a strange quark, thus enhancing relative yield of hadrons containing a combination of $cs$ (anti)quarks, e.g., $c\bar{s}$, or $csq$. Charm quark in most cases decays weakly into a strange quark (by emitting a $W^+$) hence creating a multistrange particle. 
On an example of $\Xi^-(ssd)$, we show that already for $N_{c\bar{c}}\simeq30$ (15 $c\bar{c}$ pairs), charm feed-down contribution to the yield of $\Xi^-$ is of the same order of magnitude as the experimental uncertainty of the yield measurement, in this case 7\%. Therefore, even a very conservative estimate of charm abundance at hadronization makes a significant contribution to the yield of multistrange particles, namely $\Xi(ssq),\phi(s\bar{s})$, and $\Omega(sss)$.

We also evaluated the physical bulk properties in the presence of charm and we observed a slow decrease as $N_{c\bar{c}}$ increases, with charm decays serving as an extra source of particles replacing direct contribution. However, the magnitude of this decrease is found to be insignificant compared to other sources of uncertainties of the fireball physical properties. We have checked that including charm in the SHM calculation does not contradict our previously established universality of hadronization conditions.

Without availability of experimental charm yield data, we used a fake charm hadron data point to ensure proper program operation and to test $N_{c\bar{c}}$ sensitivity. With $dN_{D^0}/dy=5\pm1$, SHARE with CHARM successfully converges to a value of $N_{c\bar{c}}\simeq 21.5$ for various initial parameter values, which implies stable behavior. We further explore the $D^0$ yield sensitivity to the charm freeze-out temperature $T_{\mathrm{charm}}$ and the total charm $N_{c\bar{c}}$ in Figure~\ref{fig:D03d}. We can see that the higher hadronization temperature is at a  given $N_{c\bar{c}}$, the lower is the yield of $D^0$. This is so since the relative abundance of heavier multicharm hadrons increases with an increase in $T_{\mathrm{charm}}$.

We believe that the new tool, SHARE with CHARM, offers comprehensive analysis of hadron production and properties of quark-gluon plasma at hadronization in contemporary heavy-ion collision experiments including the forthcoming LHC upgrade. It offers an easy to use, yet very flexible implementation of the statistical hadronization model including $u,d,s$ and now $c$ quark content. We expect SHARE with CHARM being used extensively in the coming years as charm hadron yield data from LHC should soon become available. At the time of writing this dissertation the most important result obtained with SHARE with CHARM is that universality of hadronization conditions is not affected by the charm feed into hadron abundances.

%% file: 02_Summary.tex
\chapter{SUMMARY AND FUTURE PLANS}
\label{sec:outlook}

In research leading to this dissertation, we studied particle production from quark-gluon plasma hadronization within the statistical hadronization model. Multistrange particle ratios, such as $\Xi/\phi$, are effective probes of the quark chemistry of the hadronizing fireball imprinted in the measured hadron yields. SHM is confirmed by that $\Xi/\phi$ ratio is a constant in many heavy-ion collision experiments for a wide range of collision energy and centrality. Variation of other ratios we showed implies different level of strangeness equilibration achieved during the fireball expansion, thus confirming chemical non-equilibrium variant of SHM as the only viable model of hadron production in relativistic heavy-ion collisions.

Fitting data from RHIC 62 within the statistical hadronization model framework with $\gamma_s$ as a free parameter reveals that strangeness phase space is overpopulated in most central collisions. Due to the large uncertainties of hadron data we cannot exclude the semi-equilibrium model (fixed $\gamma_q=1$) based alone on the fit quality, even though the fit clearly prefers the non-equilibrium, where for all centralities, $\gamma_q$ converges to a value near condensation of $\pi^0$, $\gamma_q\simeq1.6$, and the fireball hadronizes at a temperature of $T=140\,\mathrm{MeV}$. However, the non-equilibrium SHM approach results in the same hadronization conditions as previously observed at high energy SPS collision, suggesting both universal hadronization conditions of QGP and applicability of non-equilibrium SHM.

Analysis of preliminary data from Pb--Pb collisions at $\sqrt{s_{NN}}=2.76\,\mathrm{TeV}$ reported by the ALICE experiment at LHC confirms the hypothesis of universal hadronization conditions. We find that at the 45 times higher energy than the previously studied system at RHIC 62, the quark-gluon plasma created in heavy-ion collisions freezes-out at the same universal hadronization conditions. This result holds also as a function of collision centrality.

Detailed analysis of the full set of data shows that at LHC energy, the experimental $p/\pi$ ratio can be described by our non-equilibrium SHM variant. LHC data is precise enough to confirm the necessity of including the light quark phase space occupancy $\gamma_q > 1$ as a model parameter. The measured ratio of $p/\pi$ is a natural outcome of the fit with $\gamma_q\simeq 1.6$. No other consistent explanation has been presented up to this date; when applied consistently, posthadronization interactions do not explain all hadron species, nor do they result in a correct description of the $p/\pi$ ratio as a function of centrality.

Non-equilibrium SHM describes all available data from Pb--Pb collisions at LHC energy with high accuracy across all reported centralities, spanning 5 orders of magnitude of hadron yields at LHC alone. We confirm the universal hadronization conditions of the QGP fireball. For the most central collisions, we find a hadronization temperature of $T=138\,\mathrm{MeV}$, lower than the hadronization temperature at RHIC~62, $T=140\,\mathrm{MeV}$, for central collisions. We attribute the decrease to more supercooling of the fireball due to the more dynamical expansion, as the higher energy content implies a longer-lived fireball, which also gives enough time for strangeness to reach a steady level of QGP equilibration. This is reflected among hadrons by a saturation of $\gamma_s\simeq 2$ for wide range of centrality. Strangeness conservation during a sudden hadronization allows us to understand better the chemical equilibrium conditions of strangeness in the QGP phase. Matching the QGP Fermi gas strangeness distribution with parameters resulting from the fit to hadron data with the observed strangeness in the hadrons, we conclude that the strange quark either acquires an effective mass of $m_s =299\,\mathrm{MeV}/c^2$, or the strangeness is undersaturated in the QGP with $\gamma_s^{\small QGP}\simeq 0.77$. These bulk QGP properties require further investigation.

At LHC energy, it is estimated that a large amount of charm is produced. The $D^0$ $p_\perp$-spectrum suggests the total amount of $c+\bar{c}$ quarks ranges between $10 < N_{c\bar{c}} < 60$ $c+\bar{c}$. Scaling of the charm production cross section measured in $pp$ collisions suggest a much higher number, $100 < N_{c\bar{c}} < 400$. Since charm is predominantly produced in the initial hard parton scattering processes overpopulating the thermal phase space significantly, annihilation of $c\bar{c}$ pairs can be expected during the fireball expansion making the two estimates compatible, a point in need of future theoretical investigation. The upgraded program, SHARE with CHARM, is designed to study the effect of charm abundance on hadronization conditions in a range covering the two estimates mentioned above and beyond. 

At RHIC 200 GeV, the reported yield of charm is $N_{c\bar{c}}=c+\bar{c}=8.6$ and charm effects on hadronization remain  hidden within the large experimental hadron yield uncertainties. Without any charm hadron yield measured in Pb--Pb collisions at LHC, we can study today for each centrality, which range of $N_{c\bar{c}}$ is compatible with the non-charm data we have. Charm decay feed-down is found to be a weak constraint for $N_{c\bar{c}}$. We observe, however, that charm decays are a non-negligible source of multistrange hadrons, such as $\Xi$ or $\phi$. The contribution of charm decays in the total yield of multistrange hadrons is comparable at LHC to the small experimental uncertainty already for rather small amount of charm, $N_{c\bar{c}}\simeq 30$.

The SHM numerical computation program, SHARE with CHARM, needs as experimental input a charm hadron invariant yield to determine the amount of charm present at hadronization. With more charm hadron yield data, we can determine if charm hadronizes at the same or a different temperature $T_{\mathrm{charm}}$ than the other flavors. We made SHARE with CHARM publicly available expecting it to become a useful tool in the future to the heavy-ion community, once charm data become available.

Several future research directions have become possible as a result of this dissertation and/or due to data becoming available. I intend to address the following questions in the near future:
\begin{description}
\item[Beam Energy Scan at RHIC]\ 
To complete the study of the universal hadronization conditions of the quark-gluon plasma, we intend to study the particle production data from the Beam Energy Scan experimental program at RHIC. Detailed analysis of this data within the non-equilibrium SHM framework will test the universal hadronization conditions hypothesis. We hope to identify a reaction energy threshold for QGP creation, i.e., deconfinement.
\item[Charm dynamics in QGP fireball expansion]\
The estimated amount of charm created in the initial stages of heavy-ion collisions at LHC is well above the charm content of the fireball at hadronization suggested by the $D^0$ data. Resolution of this discrepancy between preliminary results requires a detailed modeling of charm evolution during the fireball expansion.
\item[Entropy at LHC]\ 
Charm annihilation during the fireball expansion may be the omitted additional source of entropy at LHC. Detailed modeling of flavor dynamics during the fireball expansion will reveal how much additional entropy can be created by charm annihilation after charm is created in the initial hard parton scattering processes. This will help explaining the value of strangeness over entropy ratio $s/S$ as a function of centrality at LHC.
\item[$c,s$ -- flavor hadronization dynamics]\
Because at LHC the fireball volume is so large, strange and charm quarks could undergo complex evolution in QGP towards the end of QGP expansion and hadronization. For example, distillation phenomena of both strangeness and charm need to be considered.
\item[Flavor chemistry and flavor symmetry]\ 
The precision of the LHC heavy-ion data may enable us to study the $u$ and $d$ quark flavors separately. This will test the $u$--$d$ production symmetry in the QGP fireball. Flavor is one of the important riddles of particle physics. Only in QGP can we study deconfined flavors of the first and the second elementary particle family in a large volume. 

\end{description}

\section{List of attached publications}
\begin{itemize}
\item Appendix~\ref{apx:MultistrangeRatios} (5 pages) --- M.~Petran, J.~Rafelski, Phys.Rev. C {\bf 82} (2010) 011901, 
\doi{10.1103/PhysRevC.82.011901}

\item Appendix~\ref{apx:RHIC62} (6 pages) --- M.~Petran, J.~Letessier, V.~Petracek, J.~Rafelski, proceedings of Strangeness in Quark Matter (SQM), 18-24 September 2011, Cracow, Poland\\
Acta Phys.Polon.Supp. 5 (2012) 255-262\\
\doi{10.5506/APhysPolBSupp.5.255}

\item Appendix~\ref{apx:UniversalHadronization} (5 pages) ---  M.~Petran, J.~Rafelski, Phys. Rev. C 88, 021901(R) (2013), \doi{10.1103/PhysRevC.88.021901}

\item Appendix~\ref{apx:Alice2760} (20 pages) --- M.~Petran, J.~Letessier, V.~Petracek, J,~Rafelski, Phys. Rev. C 88, 034907 (2013),
\doi{10.1103/PhysRevC.88.034907}

\item Appendix~\ref{apx:StrangeSQM} (6 pages) --- M.~Petran, J.~Letessier, V.~Petracek, J~.Rafelski, Pending publication as proceedings of Strangeness in Quark Matter (SQM), 22-27 July 2013, Birmingham, UK in IOP Conference series\\
E-print available as: \href{http://arxiv.org/abs/1309.6382}{arXiv:1309.6382 [hep-ph]}

\item Appendix~\ref{apx:ShareManual} (39 pages) ---  M.~Petran, J.~Letessier, J.~Rafelski, G.~Torrieri, Pending publication in Computer Physics Communications\\
E-print available as: \href{http://arxiv.org/abs/1310.5108}{arXiv:1310.5108 [hep-ph]}

\item Appendix~\ref{apx:CharmSQM} (4 pages) --- M.~Petran, J.~Letessier, V.~Petracek, J.~Rafelski, Pending publication as proceedings of Strangeness in Quark Matter (SQM), 22-27 July 2013, Birmingham, UK, in IOP Conference series\\
E-print available as: \href{http://arxiv.org/abs/1310.2551}{arXiv:1310.2551 [hep-ph]}
\end{itemize}

%% file: A00_XiOverPhi.tex
\chapter{MULTISTRANGE PARTICLE PRODUCTION AND THE STATISTICAL HADRONIZATION MODEL}
\label{apx:MultistrangeRatios}
\vskip-5mm
M.~Petran, J.~Rafelski, Phys.Rev. C {\bf 82} (2010) 011901\\
\doi{10.1103/PhysRevC.82.011901}
\section*{Summary}
In this work, we have motivated the SHM method of studying the multistrange hadrons, primarily $\Xi$ and $\phi$. Their ratio $\Xi/\phi$ (as defined by  Eq.~1 of the paper) is within SHM proportional to the light quark phase space occupancy $\gamma_q$. This ratio remains constant over a wide range of energy and centrality values confirming the SHM and the value of $\gamma_q$ as being constant. Other ratios also proportional to the strangeness phase space occupancy $\gamma_s$ (such as $\Xi/\pi$) vary, as can be expected, from system to system. This implies that strangeness at hadronization is out of chemical equilibrium. We must allow $\gamma_s \neq 1$ to achieve proper description of particle production in heavy-ion collisions. This result rules out the absolute chemical equilibrium among produced hadrons. Furthermore, all hadron data from Au--Au collisions at $\sqrt{s_{NN}}=62.4\,\mathrm{GeV}$ at RHIC are well described with the chemical non-equilibrium SHM. Seeing the $\Xi/\phi$ ratio being constant, we predict it to hold also for heavy-ion collisions at LHC and this prediction is now confirmed.

During the preparation of this manuscript, I was guided into SHM calculations and was responsible for the result analysis and preparation of all figures. I prepared a draft, which has been discussed and revised before submission and during the referee process.
\vfil\eject
\setlength{\textwidth}{480pt}
\setlength{\topmargin}{-35pt}
\includepdf[pages=-,pagecommand={}]{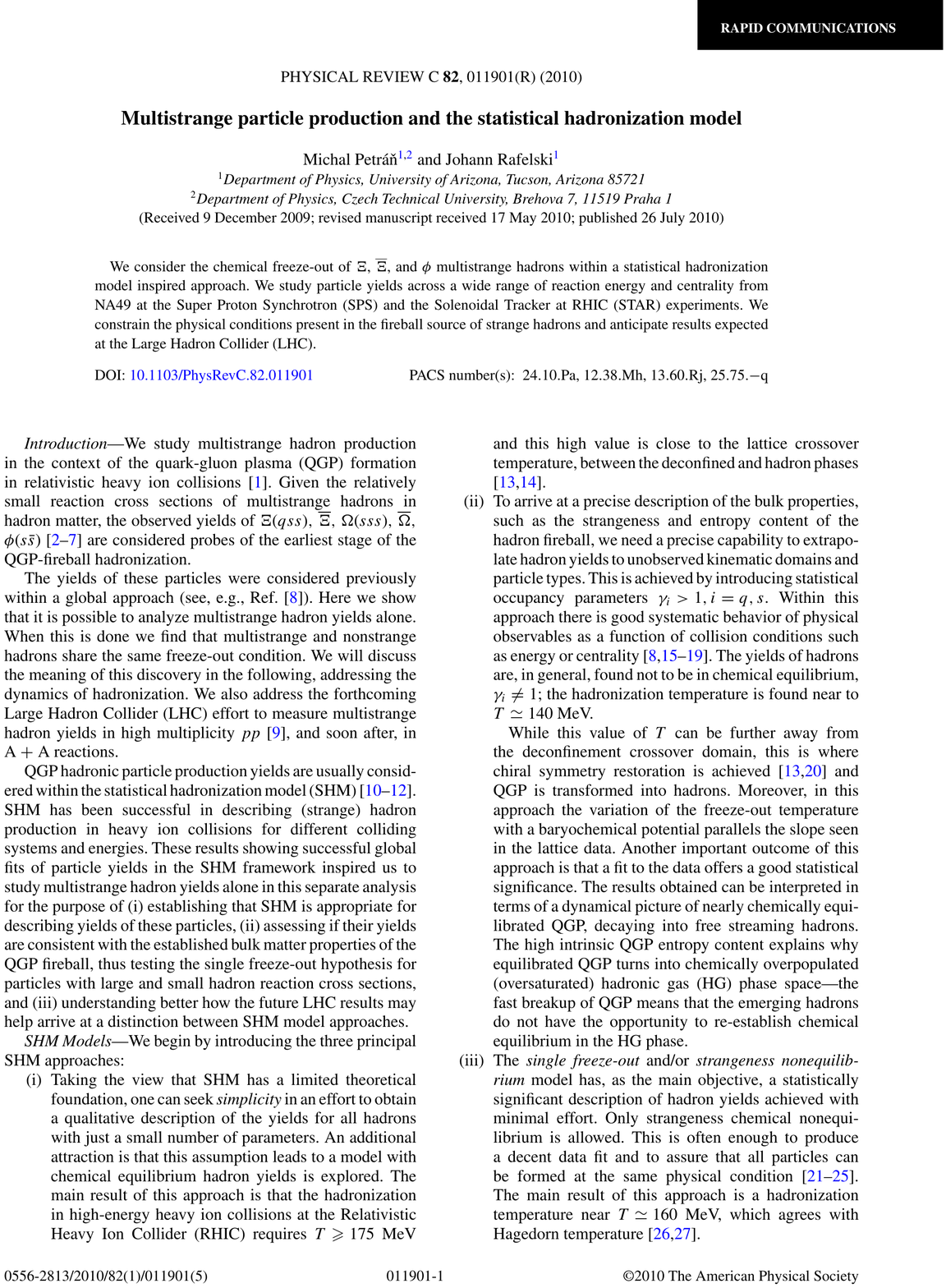}
\setlength{\textwidth}{5.9in}
\setlength{\topmargin}{0pt}

%% file: A01_RHIC62.tex
\chapter{STRANGENESS PRODUCTION IN AU--AU COLLISIONS AT $\sqrt{s_{NN}}=62.4$ GeV}
\label{apx:RHIC62}
\vskip-5mm
M.~Petran, J.~Letessier, V.~Petracek, J.~Rafelski, proceedings of Strangeness in Quark Matter (SQM), 18-24 September 2011, Cracow, Poland\\
Acta Phys.Polon.Supp. 5 (2012) 255-262\\
\doi{10.5506/APhysPolBSupp.5.255}
\vskip-2mm
\section*{Summary}

\setlength{\mylength}{\baselineskip}
\setlength{\baselineskip}{0.95\baselineskip}
We explore hadron production at RHIC in Au--Au collisions at $\sqrt{s_{NN}}=62.4\,\mathrm{GeV}$. We performed a fit to all available data as a function of centrality including multistrange particles using the chemical non-equilibrium SHM, allowing $\gamma_s \neq 1$ and $\gamma_q \neq 1$. We show that our model describes the data better (with significantly lower $\chi^2$) than previous efforts seen in literature, for instance compared to the constrained $\gamma_s<1,\gamma_q=1$ semi-equilibrium model fit presented in reference [3] of our publication. 
We find overpopulation of strangeness, $\gamma_s > 1$ for large enough system volume in central collision. We have confirmed the same hadronization conditions, pressure $P=82\,\mathrm{MeV/fm}^3$, energy density $\varepsilon=0.5\,\mathrm{GeV/fm}^3$, and entropy density $\sigma=3.3\,\mathrm{fm}^{-3}$, that were found for the high SPS energy and other RHIC energies (reference [7] of the publication). This led to a hypothesis of universal hadronization conditions in all heavy-ion collision experiments, in particular also at that time not yet available LHC. This prediction has been now confirmed.

After presenting these results at Strangeness in Quark Matter 2011 conference, I prepared a draft, which was then revised with my co-authors.
\setlength{\baselineskip}{\mylength}
\vfil\eject
\setlength{\textwidth}{480pt}
\setlength{\topmargin}{-35pt}
\includepdf[pages=-,pagecommand={}]{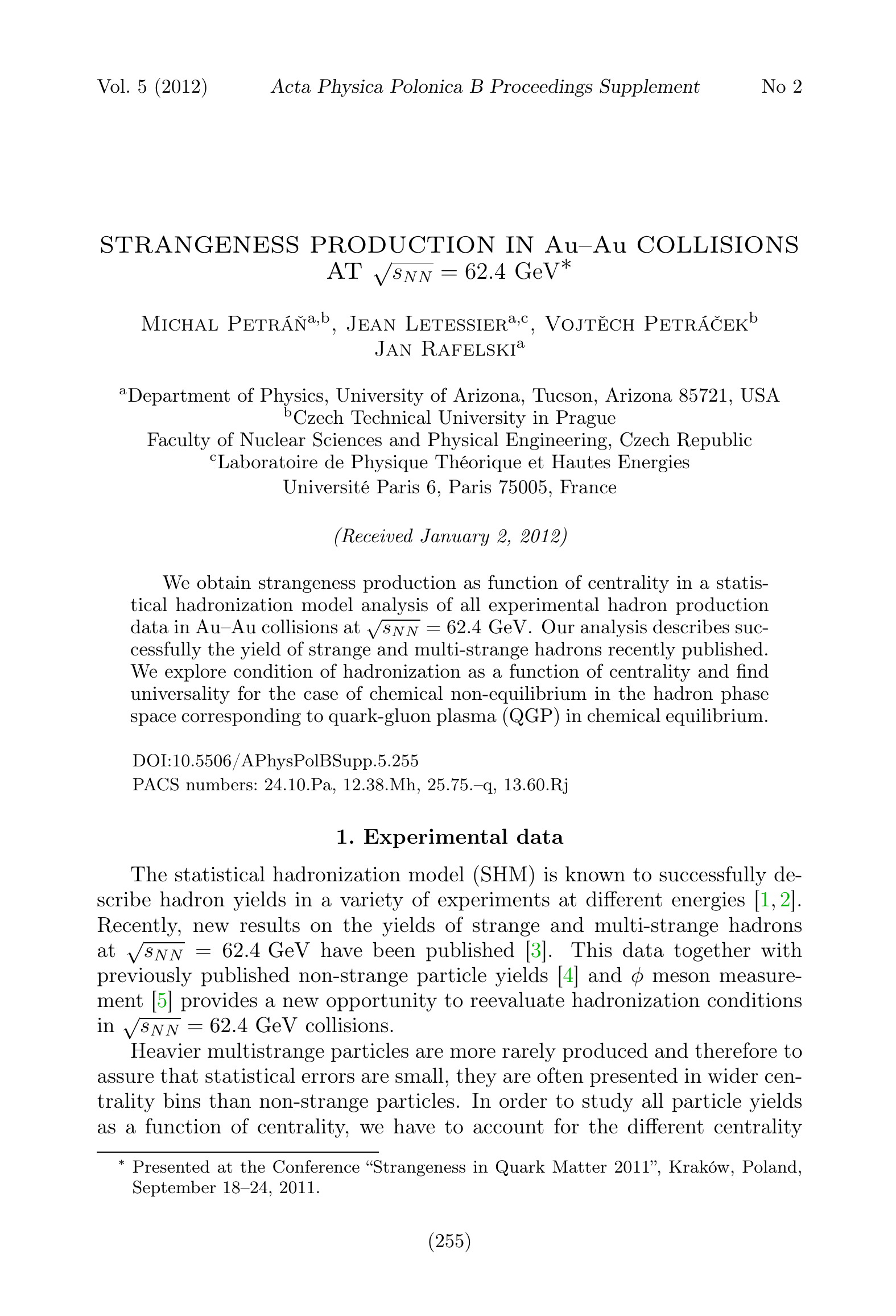}
\setlength{\textwidth}{5.9in}
\setlength{\topmargin}{0pt}

%% file: A02_UniversalHadronization.tex
\chapter{UNIVERSAL HADRONIZATION CONDITION IN HEAVY ION COLLISIONS AT $\sqrt{s_{NN}}=62$ GeV AND AT $\sqrt{s_{NN}}=2.76$ TeV}
\label{apx:UniversalHadronization}
\vskip-9mm
M.~Petran, J.~Rafelski, Phys. Rev. C 88, 021901(R) (2013)\\
\doi{10.1103/PhysRevC.88.021901}
\section*{Summary}

\setlength{\mylength}{\baselineskip}
\setlength{\baselineskip}{0.93\baselineskip}
As the first data from Pb--Pb collisions at $\sqrt{s_{NN}}=2.76\,\mathrm{TeV}$ from LHC has become available, we study the first results in four overlapping centrality bins. The first important result is that the non-equilibrium SHM describes all available data from LHC with high accuracy despite smaller data errors compared to RHIC. We estimate small baryochemical potential $\mu_B \simeq 1.5\,\mathrm{MeV}$. Following the model-independent approach presented in Appendix~\ref{apx:MultistrangeRatios}, we recall that $p/\pi \propto \gamma_q$. The experimental value of $p/\pi$ is a natural outcome of a fit with $\gamma_q\simeq 1.6$. For the first time, the data is precise enough to exclude directly SHM variants that require light quarks ($q=u,d$) in equilibrium, i.e. $\gamma_q=1$. 
Moreover, our fits to the data show the same bulk physical properties of the hadronizing fireball as at RHIC 62, despite the 45 times higher energy and 4 times larger hadronization volume in most central collisions. In other words, we establish universal hadronization conditions of QGP in all contemporary heavy-ion collision experiments. This means that bulk critical pressure $P$, energy density $\varepsilon$, and entropy density $\sigma$ are found to be in essence the same.

I was responsible for experimental data assembly, all calculations, evaluation of results and figures for this publication. I prepared the original draft of the publication, which was later revised as a response to the referee reports before it was accepted for publication. 
\setlength{\baselineskip}{\mylength}
\vfil\eject
\setlength{\textwidth}{480pt}
\setlength{\topmargin}{-35pt}
\includepdf[pages=-,pagecommand={}]{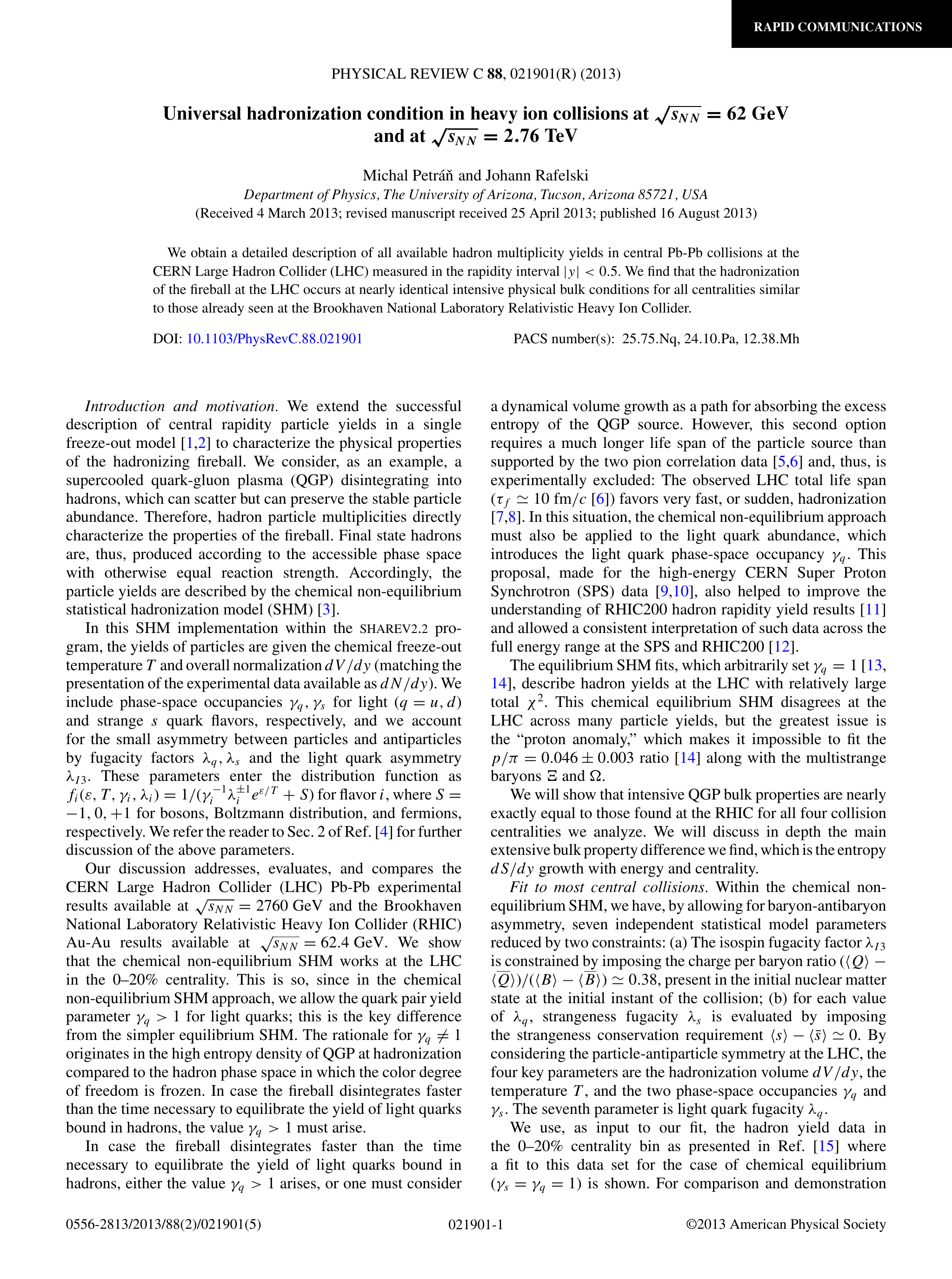}
\setlength{\textwidth}{5.9in}
\setlength{\topmargin}{0pt}

%% file: A03_Alice2.76TeV.tex
\chapter{HADRON PRODUCTION AND QUARK--GLUON PLASMA HADRONIZATION IN PB--PB COLLISIONS AT $\sqrt{s_{NN}}=2.76$ TeV}
\label{apx:Alice2760}
M.~Petran, J.~Letessier, V.~Petracek, J,~Rafelski, Phys. Rev. C 88, 034907 (2013)\\
\doi{10.1103/PhysRevC.88.034907}
\section*{Summary}
This paper extends the analysis of hadron production data in Pb--Pb collisions at $\sqrt{s_{NN}}=2.76\,\mathrm{TeV}$ obtained by the ALICE experiment at LHC. We include multistrange baryons $\Xi$ and $\Omega$ to the data set. Using interpolation of experimental results, we present a finer centrality binning of all available particle yields with non-equilibrium SHM. For comparison purposes, the semi-equilibrium and equilibrium SHM fits are also shown. We confirm the conclusion of Appendix~\ref{apx:UniversalHadronization}, that $\gamma_q > 1$ is necessary in order to describe all the measured yields within the experimental uncertainty. 

We address in more detail the $p/\pi$ ratio, which is naturally fitted within our chemical non-equilibrium SHM. A mechanism incorporating posthadronization hadron interactions, namely proton-antiproton annihilation has been proposed (see publication references [30,42--46]). We show that posthadronization interactions alter yields of all other particles and while describing the proton yield, they create large discrepancy in the multistrange baryon yields. Moreover, these interactions produce centrality dependent $p/\pi$, whereas the experimental $p/\pi$ ratio is nearly constant as a function of centrality.

We further report that at LHC, the expanding QGP reaches a steady equilibrium level of both light and strange quark flavors. This follows given the constant light quark phase space occupancy $\gamma_q\simeq 1.6$ for all centralities and the strangeness phase space occupancy $\gamma_s$, which even for relatively small systems created in peripheral collisions has a value of $\gamma_s\simeq 2$. This $\gamma_s$ value is 20\% below expectation based on the RHIC data analysis (reference [41] of the publication). This is why we found in Appendix~\ref{apx:StrangeSQM} $\gamma_s^{\small QGP}\sim 0.8$. This also contributes to lower than expected strangeness over entropy ratio $s/S\simeq 0.030$. We also report chemical freeze-out temperature $T=138\,\mathrm{MeV}$, $2\,\mathrm{MeV}$ lower compared to similar centrality at RHIC. We interpret this as a supercooling of the more dynamically expanding QGP fireball. However, the difference is within error of RHIC results.

Some of the data we used in the analysis were still preliminary. During the review process of this paper, final yields of $K_S^0, \Lambda, \Xi$ and $\Omega$ were published (references [61,62] of the publication). We have confirmed (see section IV.F: Update of our publication) that the final centrality dependent data is compatible with all our results in the paper.

Preparation of this article took a considerable amount of time. I was responsible for the data collection and calculations, I analyzed the results and prepared all figures for publication and an original manuscript draft. The text was then extended several times by all co-authors including added results before and one revision during the referee review process.

\vfil\eject
\setlength{\textwidth}{480pt}
\setlength{\topmargin}{-35pt}
\includepdf[pages=-,pagecommand={}]{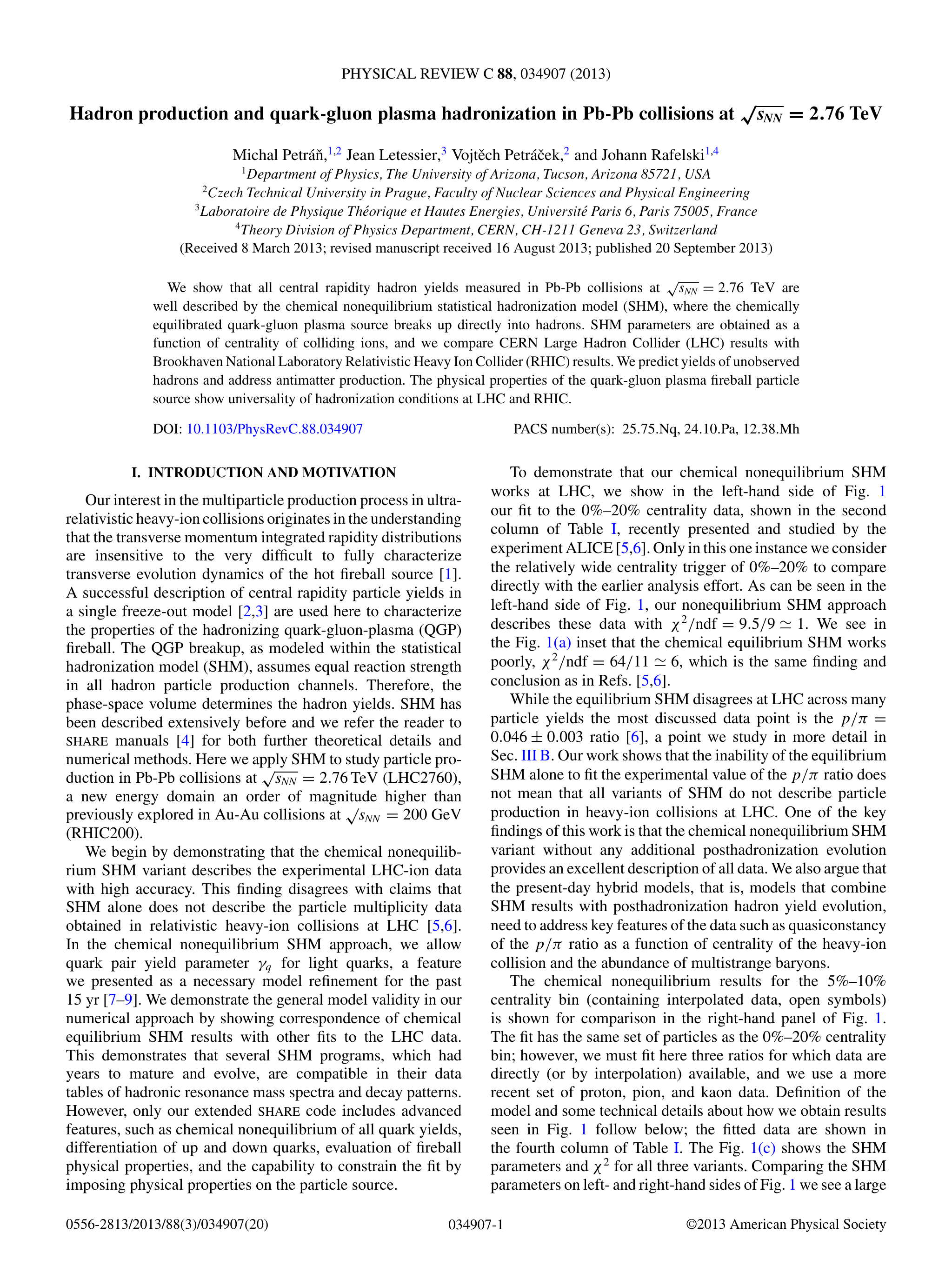}
\setlength{\textwidth}{5.9in}
\setlength{\topmargin}{0pt}

%% file: A04_SQM2013-strangeness.tex
\chapter{INTERPRETATION OF STRANGE HADRON PRODUCTION AT LHC}
\label{apx:StrangeSQM}
M.~Petran, J.~Letessier, V.~Petracek, J~.Rafelski, Pending publication as proceedings of Strangeness in Quark Matter (SQM), 22-27 July 2013, Birmingham, UK, in IOP Conference series\\
E-print available as: \href{http://arxiv.org/abs/1309.6382}{arXiv:1309.6382 [hep-ph]}
\section*{Summary}
This publication studies the physical stability of the fit presented in Appendix~\ref{apx:Alice2760} in a representative centrality bin 10--20\%, for which no data interpolation is necessary. We reproduce the fit from Appendix~\ref{apx:Alice2760} with the updated data set. We then run a fit with finite resonance widths, which converges to the same minimum in terms of model parameters and physical bulk properties. This gives us confidence, that we can omit this computationally very intensive feature in our calculations.

The yield of $\Lambda$ is identified as the main contributor to the total fit $\chi^2$; its model yield is about 1.1 standard deviation below experimental value. However, the model parameters are constrained enough by the rest of the experimental data set, so that the $\Lambda$ data point does not affect the parameter values, but only increases the total $\chi^2$ of the fit. The hadron spectrum in our SHM program includes only 3-star (***) and 4-star (****) states from the Review of Particle Physics (publication reference [5]). We explore the influence of an omitted 2-star (**) hadron resonance, $\Sigma(1560)$, as an additional source of $\Lambda$. When we add $\Sigma(1560)\to\Lambda\pi$ to the particle list and decay tree, the fit converges to the same minimum, and the $\Lambda$ yield is now fitted within $^1\!/\;\!\!_2$~standard deviation of its experimental value. This finding has prompted an experimental search for this resonance.

We study how well is strangeness conserved at hadronization. From the measured strange hadron yields we know that over 600 strange and antistrange quarks are produced per unit rapidity at hadronization in the most central Pb--Pb collisions. We calculate the total strangeness (i.e. the number of strange and antistrange quarks $s+\bar{s}$) in the QGP phase using the Fermi gas strangeness density and match it to the measured strangeness in hadrons.  Assuming the sudden hadronization implying strangeness conservation at hadronization, we report two scenarios: 1) strangeness in the QGP fireball is in equilibrium and the strange quark acquires an effective mass of $m_s = 299\,\mathrm{MeV}/c^2$ at hadronization; or 2) the strange quark has its PDG mass of $m_s = 140\,\mathrm{MeV}/c^2$ (at the relevant scale $2\pi T\simeq 0.9$ GeV) implying the QGP fireball strangeness to be undersaturated with strangeness phase space occupancy in the plasma $\gamma^{\small{QGP}}_s=0.77$.

After presenting these results in a plenary lecture Strangeness in Quark Matter conference, I prepared a draft for the written report. I was responsible for all calculations, result analysis and figure preparation. My co-authors and I revised the text before submission.

\vfil\eject
\setlength{\textwidth}{450pt}
\setlength{\topmargin}{-15pt}
\includepdf[pages=-,pagecommand={}]{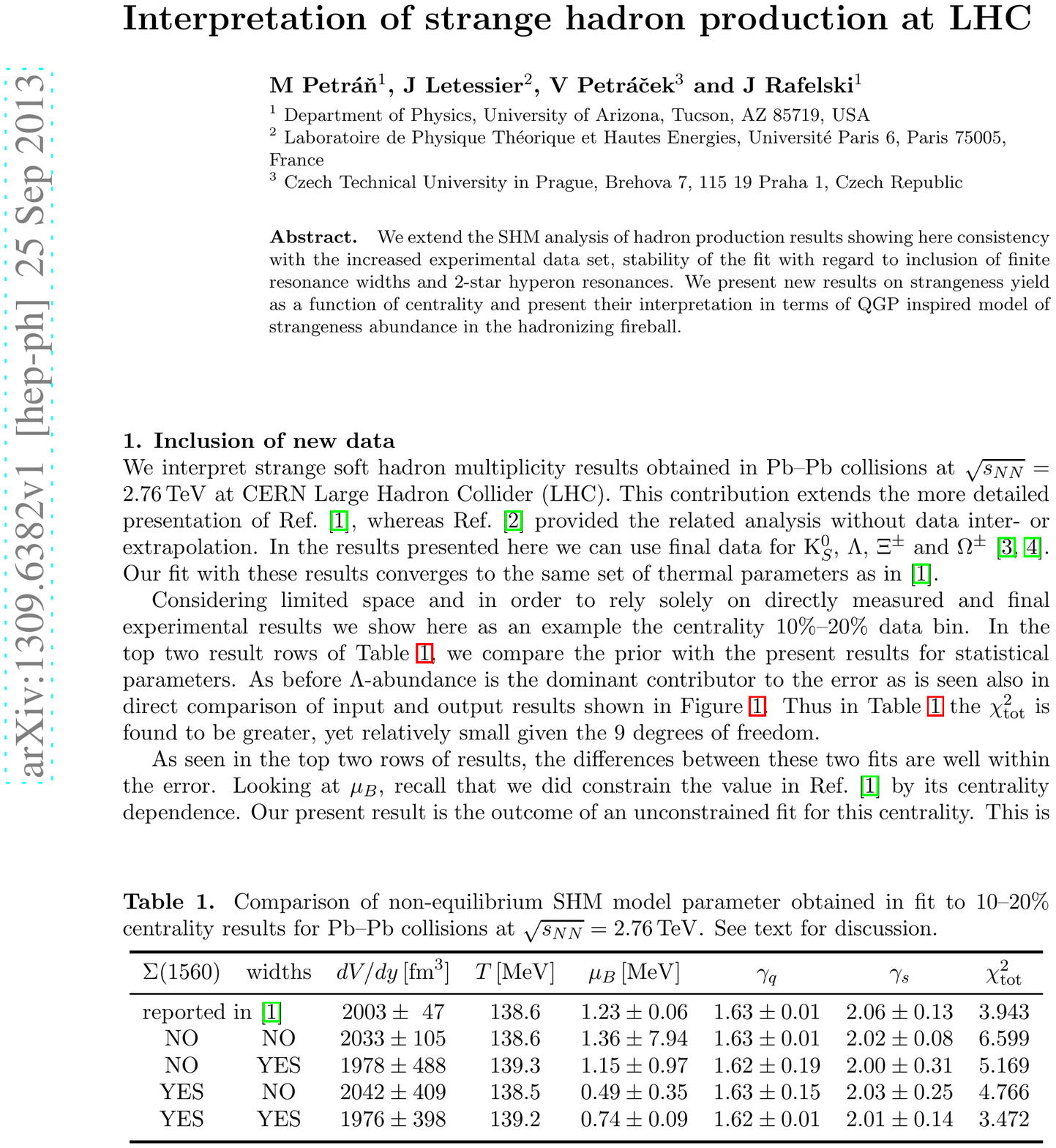}
\setlength{\textwidth}{5.9in}
\setlength{\topmargin}{0pt}

%% file: A06_Share3Manual.tex
\chapter{SHARE WITH CHARM}
\label{apx:ShareManual}
M.~Petran, J.~Letessier, J.~Rafelski, G.~Torrieri, Pending publication in Computer Physics Communications\\
E-print available as: \href{http://arxiv.org/abs/1310.5108}{arXiv:1310.5108 [hep-ph]}
\section*{Summary}
The predicted charm production at the top LHC heavy-ion collisions called for an upgrade of the SHM implementation, SHAREv2. I upgraded SHARE with a CHARM module, hence SHARE with CHARM, included all charmed 3-star (***) and 4-star (****) hadron states from the current PDG review (publication reference [4]) together with their numerous decay channels. Frequently, the experimentally measured charm decays today have considerable uncertainty, which presents an intrinsic uncertainty in all our results with charm. We also included all decay channels that are suggested by, e.g., the isospin symmetry. assuming that all branching ratios add up to 1.

The new version of the SHARE program adds several new functionalities and provides a common SHM framework for all contemporary heavy-ion experiments. The default settings of the program are tuned for the LHC environment; for example, the omission of all weak decay feed-down, which is possible given enhanced tracking capabilities of the ALICE experiment. We implement the number of charm and anticharm quarks $N_{c\bar{c}} = c+\bar{c}$ as a new model parameter as well as the ratio of charm to light hadron freeze-out temperature $T_{charm}/T$, which allows more flexibility in study of charm hadronization. When an invariant yield of a charm hadron becomes available, SHARE with CHARM provides the total amount of charm, predicts other charm hadron yields, evaluates any change in bulk physical properties due to charm, and ultimately demonstrates the hypothesis that the statistical hadronization model describes the production of charm hadrons in QGP hadronization.

The publication is a complete detailed user's manual to the SHARE with CHARM that includes description of features introduced in SHAREv1 and SHAREv2 as well as the new capabilities introduced in this release, combined into one concise user's guide.

For the program upgrade, it was necessary to collect charm hadron properties and decay channels, which I compiled into  program input files, solving inconsistencies with my co-authors. I was responsible for the CHARM module coding and necessary code modifications of the original SHAREv2 program. I prepared the first draft of the program manual, which was revised together with all co-authors.

\vfil\eject
\setlength{\textwidth}{450pt}
\setlength{\topmargin}{-15pt}
\includepdf[pages=-,pagecommand={}]{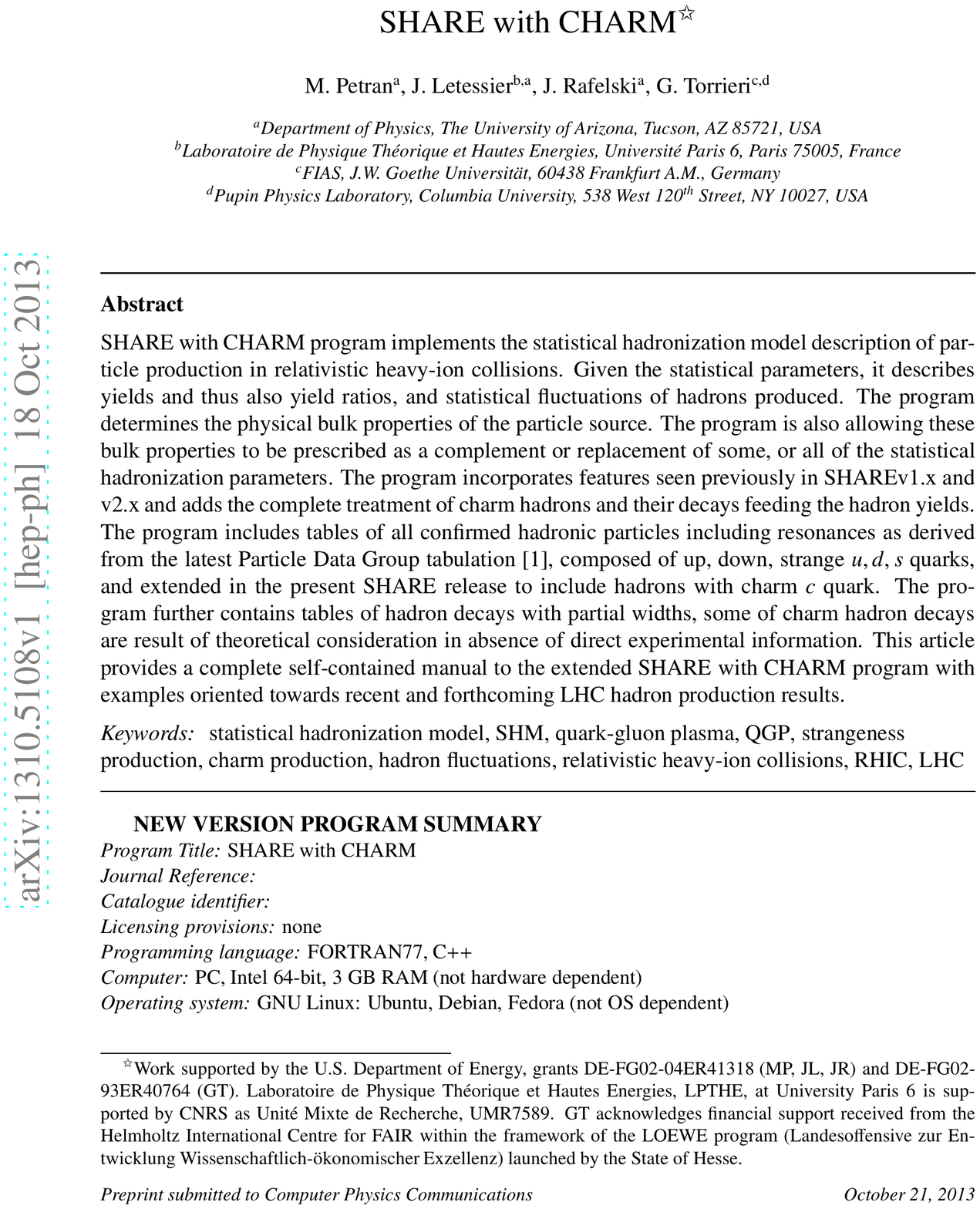}
\setlength{\textwidth}{5.9in}
\setlength{\topmargin}{0pt}

%% file: A05_SQM2013-charm.tex
\chapter{CHARM DECAY AS A SOURCE OF MULTISTRANGE PARTICLES}
\label{apx:CharmSQM}
M.~Petran, J.~Letessier, V.~Petracek, J.~Rafelski, Pending publication as proceedings of Strangeness in Quark Matter (SQM), 22-27 July 2013, Birmingham, UK, in IOP Conference series\\
E-print available as: \href{http://arxiv.org/abs/1310.2551}{arXiv:1310.2551 [hep-ph]}

\section*{Summary}
We evaluate the effect of including charm flavor hadrons produced in heavy-ion collisions at RHIC and LHC. Based on charm cross section measurement (publication reference [1]), we estimate the amount of charm produced in central Au--Au collisions at the top RHIC energy 200 GeV to be $N_{c\bar{c}}=8.6$ at mid-rapidity. Our study shows that charm effect at RHIC is negligible in terms of the resulting influence on SHM parameters and bulk physical properties of the source. However, at LHC the expected amount of charm is much greater. Theoretical charm cross section prediction leads to $N_{c\bar{c}}=246\pm154$ at mid-rapidity in central Pb--Pb collisions (publication reference [2]). We complement this estimate with the incomplete $p_\perp$-spectrum of $D^0$ meson measured by the ALICE collaboration (publication reference [7]), which leads to $dN_{D^0}/dy \in (1.3,9.0)$. The $D^0$ yield estimate translates to a range of $N_{c\bar{c}} \in (6,45)$ in 0--20\% centrality Pb--Pb collisions at $\sqrt{s_{NN}}=2.76\,\mathrm{TeV}$, much lower than the theoretical expectation. We evaluate the charm effect in a range of $N_{c\bar{c}}$ covering both scenarios. We confirm a previous result (publication reference [6]), that charm decays are a significant source of multistrange particles and we show that for $N_{c\bar{c}}>50$, the charm effect on the yield of $\Xi$, for example, is larger than experimental error and hence charm is relevant to achieve precise description of particle production at LHC. We show that fitted charm hadron decays replace strange hadron production from QGP and effectively reduce the strangeness phase space occupancy $\gamma_s$ as a function of charm abundance. Physical properties of the fireball are only marginally different. 

During preparation of this proceedings, I was responsible for all calculations, result analysis and preparation of graphic presentation. My original draft was revised before submission with my co-authors.

\vfil\eject
\setlength{\textwidth}{450pt}
\setlength{\topmargin}{-15pt}
\includepdf[pages=-,pagecommand={}]{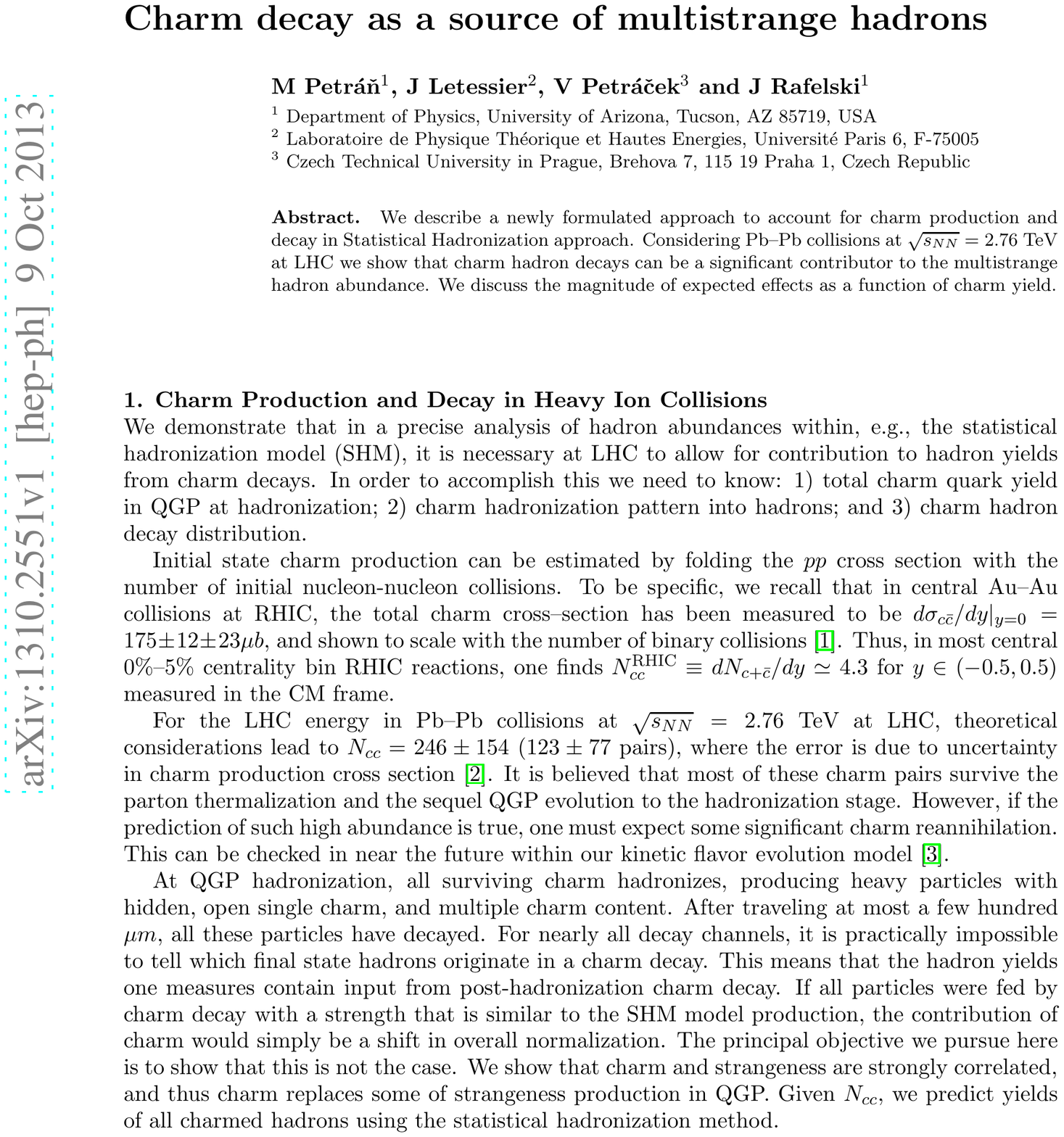}
\setlength{\textwidth}{5.9in}
\setlength{\topmargin}{0pt}